\documentclass[10pt,twocolumn,compsoc]{IEEEtran}

\usepackage{cite}
\usepackage[pdftex]{graphicx}
\usepackage{amsmath}
\usepackage{algorithmic}
\usepackage{array}
\usepackage{fixltx2e}
\usepackage{enumitem}
\usepackage{balance}
\usepackage{color, colortbl}
\usepackage{longtable}
\usepackage{wrapfig}
\usepackage{tcolorbox}
\newtcolorbox{blockquote}{colback=gray!5,boxrule=0.4pt,colframe=black,fonttitle=\bfseries,top=2pt,bottom=2pt}

\usepackage{tabularx}
\usepackage{multirow}
\usepackage{color,colortbl}
\definecolor{lightgray}{gray}{0.8}
\usepackage{adjustbox}
\usepackage{hyperref}

\usepackage{amssymb}
\usepackage{pifont}
\newcommand{\cmark}{\ding{52}}
\newcommand{\xmark}{\ding{56}}

\newcommand{\respto}[1]{}

\newcommand{\revised}{\textcolor{black}}

\ifCLASSOPTIONcompsoc
    \usepackage[caption=false,font=normalsize,labelfont=sf,textfont=sf]{subfig}
\else
    \usepackage[caption=false,font=footnotesize]{subfig}
\fi

\usepackage{dblfloatfix}

\newcommand{\quart}[4]{\begin{adjustbox}{max width=.1\textwidth}\begin{picture}(100,5)
    {\color{black}\put(#3,2){\circle*{7}}\put(#1,2){\line(1,0){#2}}}\end{picture}\end{adjustbox}}

\newcommand{\fig}[1]{Figure~\ref{fig:#1}}
\newcommand{\tbl}[1]{Table~\ref{tbl:#1}}

\newcommand{\bi}{\begin{itemize}}
\newcommand{\ei}{\end{itemize}}

\begin{document}

\title{How Different is Test  Case Prioritization    \\ for Open and Closed Source Projects?  }

\author{Xiao~Ling,
        Rishabh~Agrawal,
        and~Tim~Menzies,~\IEEEmembership{Fellow,~IEEE}
\IEEEcompsocitemizethanks{\IEEEcompsocthanksitem X. Ling, R. Agrawal and  T. Menzies are with the Department of Computer Science, North Carolina State University, Raleigh, USA.
\protect 
E-mail: lingxiaohzsz3ban@gmail.com,  ragrawa3@ncsu.edu,  timm@ieee.org}}

\markboth{IEEE Transactions on Software Engineering}%
{Ling \MarkLowerCase{\textit{et al.}}: How Different is Test Case Prioritization for Open and s for IEEE Journals}
 

\IEEEtitleabstractindextext{
\begin{abstract}
Improved test case prioritization means that   software developers can detect and fix more software faults sooner than usual.
But is there one ``best'' prioritization algorithm? Or do different kinds of projects deserve special kinds of prioritization?
To answer these questions, this paper applies nine prioritization schemes to 31 projects that range from
(a)~highly rated open-source Github projects to
(b)~computational science software to   (c)~a closed-source project. 
We find that prioritization approaches that work best for open-source projects can work worst for the closed-source project (and vice versa).
From these experiments, we conclude that 
(a) it is ill-advised to always
apply one prioritization  scheme to all projects
since (b) prioritization requires tuning to different project types.

\end{abstract}

}

\maketitle

\section{Introduction} \label{Introduction}
\IEEEPARstart{R}{egression testing}
is widely applied in both open-source projects and closed-source projects~\cite{fazlalizadeh2009prioritizing, lu2009introReg, mahajan2015intorReg}.
When software comes with a large regression suite, then developers can check if their new changes damage old functionality.

Excessive use of regression testing can be expensive and time-consuming,
especially if run after each modification to the software. Such high-frequency regression testing can consume as much as 80 percent of the testing budget and require half the software maintenance effort~\cite{chittimalli2007cost}.

To reduce the cost of performing regression testing, {\em test case prioritization (TCP)} is widely studied in software testing~\cite{beller2019IDE, beller2015and, fazlalizadeh2009prioritizing, lu2009introReg, Wong1997TCPwidelyapply, cho2016history, elbaum2014techniques}. 
\revised{In test case prioritization}, some features are extracted from prior test suites and test results and then applied to prioritize 
\revised{test cases in the current round}.
Google reports that test case prioritization can reduce the time for programmers to find 50\% of the failing tests from two weeks to one hour~\cite{elbaum2014techniques}.

Prioritization schemes that work on some projects may fail on others. 
As shown later in this paper, not all projects track the information required for all the different prioritization algorithms. For example, suppose closed-source projects are prioritized using the textual descriptions of the test cases. That approach may not always work for open-source projects where such textual descriptions may be absent.
Previously, Yu et al.~\cite{yu2019terminator} reported that the TERMINATOR test case prioritization algorithm was better than dozens of alternatives. However, TERMINATOR was developed for closed-source proprietary software. This raises the question: does TERMINATOR work for other kinds of projects (e.g. open-source projects)? 

To explore this issue, this paper applies \revised{test case prioritization} schemes to data from a closed-source proprietary project and 30 open-source projects. To the best of our knowledge, this study explores more prioritization algorithms,
\revised{as well as}
more kinds of data than prior work. Using that data, we answer the following research questions. \newline
\textbf{RQ1: What is the best algorithm for the closed-source project?} 
We find that we can reproduce prior results:
\begin{quote}
{\em As seen before, the TERMINATOR prioritization scheme works best for that closed-source project.}
\end{quote}
\textbf{RQ2: What is the best algorithm for open-source projects?} 
While our {\bf RQ1} results concurred with past work, {\bf RQ2} shows that closed-source prioritization methods should not
be applied to open-source projects:
\begin{quote} 
{\em For open-source projects, the best approach is not TERMINATOR, but rather to prioritize  using either passing times since the last failure or another exponential metric (defined in \S\ref{TCPAlgorithms}).} 
\end{quote} 
\textbf{RQ3: Do different prioritization algorithms perform various in the open-source projects and the closed-source project?} Combining RQ1 and RQ2, we can assert:
\begin{quote}
{\em Test case prioritization schemes that work {\em best} for the industrial closed-source project can work worse for open-source projects (and vice versa).}
\end{quote}
The rest of this paper is structured as follows. Section~\ref{Background} describes related work and Section~\ref{Methods} explains our experimental methods.
 Section~\ref{Results} shows answers to the above questions. This is followed by
 some discussion in Section~\ref{Discussion} and a review of threats to validity in Section~\ref{ThreatstoValidity}, Section~\ref{ConclusionandFutureWork} shows our conclusion:
\begin{quote}
{\em It is ill-advised to always apply one prioritization scheme to all projects since prioritization requires tuning to different projects types.}
\end{quote}
To say that another way, prioritization schemes should always be re-assessed using local data. To simplify that process, we have made \revised{all the scripts and data used in this study} available on-line \footnote{https://github.com/ai-se/TCP2020}. Note that those scripts include all the major history-based prioritization schemes
seen in the current literature.

 

  
\section{Background} \label{Background}

\subsection{Definitions} \label{Definitions}
This paper shows that  the  ``best''
prioritization differs between the closed-source proprietary project and the open-source projects. These projects can be distinguished as follows:
\bi
    \item
    \textbf{Open-source projects} are developed and distributed for free redistribution, the possibility  for modifications, and with full access to the  source code~\cite{koch2002effort, raja2012defining}.
    \item
    \textbf{Closed-source projects} are proprietary software, developed with authorized users with private modification, republishing under a permission agreement~\cite{saltis2018comparing}.
    \ei
    As to the sites where we collect data:
    \bi
    \item  
    \textbf{Github}
    is a  hosting company for software development version control.
  Free GitHub accounts are commonly used to host open-source projects. As of January 2020, GitHub reports having over 40 million users and more than 100 million repositories (including at least 28 million public repositories), making it the largest host of source code in the world.

    \item
    \textbf{TravisTorrent} is a public data set \revised{which contains} vanilla API data (build information), the build log analysis (tests information), plus repository and commit data~\cite{msr17challenge}.
    
    \item
    \textbf{Travis CI} is an OSS continuous integration as-a-service platform that can run the test suits automatically after \revised{each of the commit in GitHub}~\cite{beller2017oops}. 
\ei

\begin{table*}[!t]
    \centering
    \caption{Summary of literature.  ``\#Scheme'' shows number of    prioritization methods studied \revised{in that literature}.   ``\# Closed'' and ``\# Open'' shows how much data was used (measured in terms of number of projects).}
    \label{tbl:PaperTable}
    \scriptsize
    \begin{tabular}{r@{~}rccccrr}
        \rowcolor{gray!25}\multicolumn{2}{r}{}  & \# Scheme & \# Closed & \# Open & Year & Venue & Citations \\
        \hline 
        Prioritizing Test Cases For Regression Testing&\cite{rothermel2001prioritizing} & 9 & 0 & 8 & 2001 & TSE & 1345 \\
        \rowcolor[HTML]{F3F3F3}   Test case prioritization: A family of empirical studies&\cite{elbaum2002test} & 18 & 0 & 8 & 2002 & TSE & 994 \\
        Search algorithms for regression test case prioritization&\cite{li2007search} & 5 & 0 & 6 & 2007 & TSE & 739 \\
        \rowcolor[HTML]{F3F3F3}  A history-based test prioritization technique ...&\cite{kim2002history}  & 1 & 0 & 8 & 2002 & ICSE & 461 \\
        Adaptive random test case prioritization&\cite{jiang2009adaptive} & 9 & 0 & 11 & 2009 & ASE & 222 \\
        \rowcolor[HTML]{F3F3F3}   System Test Case Prioritization of New and Regression Test Cases&\cite{srikanth2005system} & 1 & 0 & 0 & 2005 & ESEM & 223 \\
        Techniques for improving regression testing in continuous integration...&\cite{elbaum2014techniques}  & 1 & 1 & 0 & 2014 & FSE & 187 \\
        \rowcolor[HTML]{F3F3F3} A clustering approach to improving test case prioritization...&\cite{carlson2011clustering} & 1 & 1 & 0 & 2011 & ICSM & 97 \\
        Test case prioritization for black box testing&\cite{qu2007test} & 2 & 2 & 0 & 2007 & COMPSAC & 94 \\
        \rowcolor[HTML]{F3F3F3} Test case prioritization for continuous regression testing...&\cite{marijan2013TCP} & 1 & 1 & 0 & 2013 & ICSM & 87 \\
        Prioritizing test cases for resource constraint environments...&\cite{fazlalizadeh2009prioritizing} & 2 & 0 & 7 & 2009 & ICCSIT & 32 \\
        \rowcolor[HTML]{F3F3F3}   Prioritizing manual test cases in traditional \& rapid release environments&\cite{hemmati2015prioritizing} & 3 & 0 & 1 & 2015 & ICST & 30 \\
        History-based test case prioritization for failure information&\cite{cho2016history} & 1 & 0 & 2 & 2016 & APSEC & 11 \\
        \rowcolor[HTML]{F3F3F3}   Test re-prioritization in continuous testing environments&\cite{zhu2018test} & 1 & 2 & 0 & 2018 & ICSME & 10 \\\hline
  \rowcolor{gray!25}\multicolumn{2}{r}{This paper $\Longrightarrow$}  & 9 & 1 & 30 & 2020 & TSE&  
    \end{tabular}
\end{table*}

\begin{figure*}[!b]
    \centering
    \includegraphics[width=.8\textwidth]{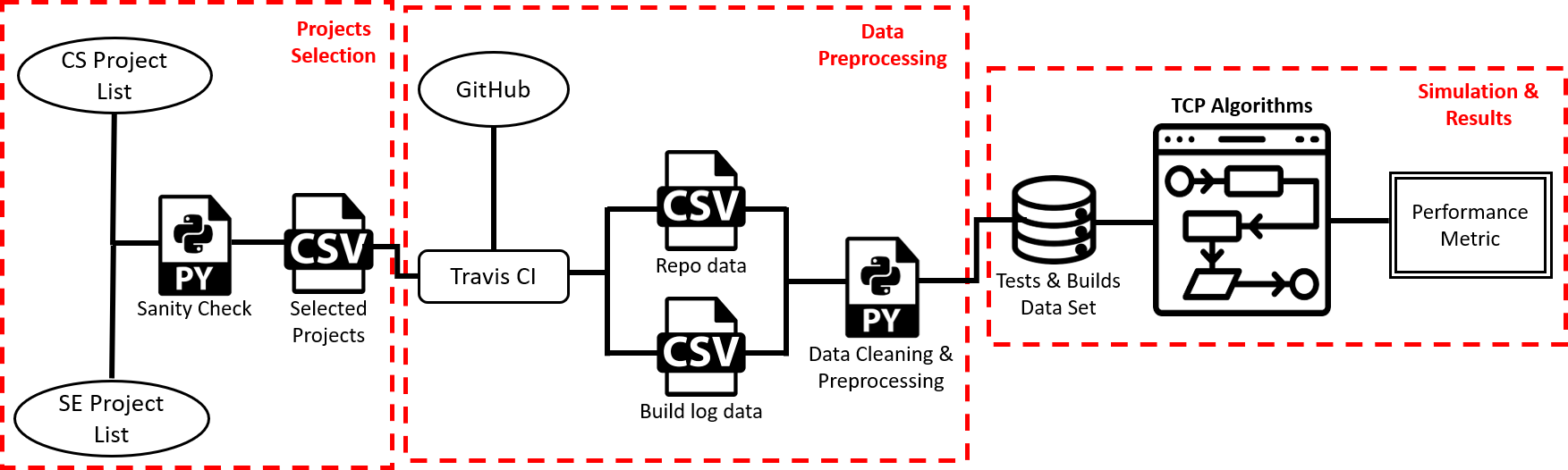}
    \caption{Framework for our Experiment}
    \label{fig:System Framework}
\end{figure*}

\subsection{Why Study Test Case Prioritization?} \label{WhyStudyTCP}
In software development, regression testing is very important in detecting software faults. However, it is also widely recognized as an expensive process. The most helpful approach to reduce computational cost and potentially reveal faults earlier is called test case prioritization~\cite{li2007search, rothermel1999test, elbaum2002test, elbaum2001incorporating, do2006use}. \revised{A} better test case prioritization \revised{scheme} is useful since:
\bi
\item When developers extend   a code base, they can check that their new work does not harm existing functionality.
\item This, in turn, enables a rapid release process where developers can safely send new versions of software to  users  each week (or even each day). 
    \item
    Faults can be revealed earlier than normal execution, which significantly increases the efficiency and reduces the cost of regression testing. Moreover, within a time limit, more faults can be detected by performing test cases prioritization~\cite{fazlalizadeh2009prioritizing, lu2009introReg, haidry2013TCP, marijan2013TCP}.
    \item
    Test managers can locate and fix faults earlier than normal execution by \revised{applying test case prioritization}~\cite{Malz2012TCP}.
\ei
There are many scenarios where the test case prioritization results of this paper can be applied. According to Zemlin~\cite{zemlin2017}, 80 percent of current software projects are open-source projects. Some projects even have extensive test suites. To maintain the stability of projects, project developers want to detect more faults in a limited time after each modification. For that purpose, test case prioritization is widely applied in regression testing~\cite{beller2019IDE, beller2015and, fazlalizadeh2009prioritizing, lu2009introReg, Wong1997TCPwidelyapply, cho2016history, elbaum2014techniques}. Therefore, a well-performed prioritization algorithm for open-source projects is highly demanded, which can let project developers:
\bi
    \item
    Detect more faults within a period of time.
    \item
    Start to fix software bugs earlier than usual.
\ei
Moreover, test case prioritization is applied in the industrial closed-source projects. For example, LexisNexis is an industrial company that provides legal research, risk management, and business research~\cite{vance2010legal}. The Lexis Advance platform is maintained by a set of automated UI tests, which is a case of regression testing. Such testing tasks are very expensive in execution time. Yu et al. state that the automated UI test suite that LexisNexis uses on testing takes \revised{approximately} 30 hours to execute~\cite{yu2019terminator}. Therefore, LexisNexis seeks a prioritization algorithm that can help developers to
\begin{enumerate}
    \item
    Test software more often, \revised{and} ship more updates to customers, at a faster rate;
    \item
    Save time when waiting for feedback on the last change~\cite{yu2019terminator}.
\end{enumerate}

\subsection{Who studies Test Case Prioritization?}
\label{WhoStudyTCP}
For all the  above reasons,
many researchers explore the  test case prioritization approach. For example:
\bi
    \item
    Yu et al. introduced an active learning-based framework called TERMINATOR, which implements a Support Vector Machine classifier to achieve higher fault detection rates on automated UI testing~\cite{yu2019terminator}.
    \item
    Hemmati et al. propose a risk-driven clustering method that assigns the highest risk to the tests that failed in the closest version before the current version. After that, tests that failed in the two versions before the current version will be assigned and so on~\cite{hemmati2015prioritizing}.
    \item
    Fazlalizadeh et al. propose a test case fault detection performance approach which calculates the ratio of the number of times that the execution of the test case fails to the number of executions of the test case~\cite{fazlalizadeh2009prioritizing}.
    \item
    Kim et al. claim that the selection probabilities of each test case at each test run is  useful in prioritization. They propose an  ``Exponential Decay Metric'' (defined later in this paper) which can calculate selection probabilities with weighted individual history observation~\cite{kim2002history}.
    \item
    Zhu et al. and Cho et al. study the correlations between two test cases. They introduce different test case prioritization approaches based on different information on correlation. Zhu et al. purpose co-failure distributions, while Cho et al. implement the flipping history of two test cases~\cite{zhu2018test, cho2016history}. Based on Cho et al., two test cases are ``flipped'' if they change to the opposite status (Pass or Fail) in two consecutive runs~\cite{cho2016history}.
    \item
    Li et al. study five search techniques (Hill Climbing, Genetic Algorithm, Greedy, Additional Greedy, and 2-Optimal Greedy) for code coverage~\cite{li2007search}.
    \item
    Elbaum et al. use four approaches with function coverage information of test cases. They point out that different testing scenarios should apply the appropriate prioritization approach~\cite{elbaum2004selecting}.
    \item
    For more examples, see~\cite{jiang2009adaptive, elbaum2001incorporating, malishevsky2006cost, Malz2012TCP, do2006use, zhang2009time, jeffrey2006test}.
\ei
\subsection{How to Study Test Case Prioritization?} \label{HowStudyTCP}
In order to base this work on current methods in the literature, we base this paper on two literature reviews of test case prioritization. In March 2019, Yu et al. explored 1033 papers by using incremental text mining tools and found a list of   prioritization algorithms that covered the 
previously outperformed test case prioritization methods
in this area~\cite{yu2019terminator}. 
To confirm and extend that finding for different types of projects, in May 2020, we conducted our own review. Beginning with papers from senior SE venues (as defined by  Google Scholar Metrics ``software systems''), we searched for highly cited or recent papers studying \textit{test case prioritization}. For our purposes, ``highly cited'' means at least 10 citations per year since publication. 
This search found a dozen high profile test prioritization papers in the last 10 years. To that list, we used our domain knowledge to add two \revised{papers} that we believed to be the most influential early contributions to this work. The final list of 14 papers is \revised{in} \tbl{PaperTable}.


Based on the papers in \tbl{PaperTable}, and the study of Yu et al., we find that the following history-based information is \revised{frequently} used in test case prioritization. We exclude coverage-based algorithms in this work because (a) collecting proper coverage information for each build in our open-source projects is not only hard but also time-consuming and (b) our proprietary project from our industrial partner is private. \revised{We cannot access the coverage information}. Note that any term {\em in italics} is defined later in this paper (see \S\ref{TCPAlgorithms}).
 
\bi
    \item
    \textit{Time since last failure}: Prioritize test cases by using the numbers of consecutive non-failure~\cite{hemmati2015prioritizing, elbaum2014techniques}.
    \item
    \textit{Failure rate}: Prioritize test cases by the ratio of total failure times over total execution times~\cite{fazlalizadeh2009prioritizing}.
    \item
    \textit{Exponential Decay Metrics}: Prioritize test cases by applying Exponential Decay Metrics, which adds weights in execution history~\cite{kim2002history}.
    \item
    \textit{ROCKET Metrics}: Prioritize test cases by applying ROCKET 
4
 Metrics~\cite{marijan2013TCP}.
    \item
    \textit{Co-failure}: Prioritize test cases by Co-failure distribution information~\cite{zhu2018test}.
    \item
    \textit{Flipping History}: Prioritize test cases by the correlations of flipping history~\cite{cho2016history}.
    \item
    \textit{TERMINATOR}: An active learning method~\cite{yu2019terminator}.
\ei


\begin{wraptable}{r}{1.4in} 
 
    \caption{Sanity Check. From~\cite{tu2020changing}}
    \label{tbl:SanityCheck}
    \scriptsize
    \begin{tabular}{l|l}
      \rowcolor{gray!25}
        {\textbf{Test}} & {\textbf{Criteria}} \\
        \hline
        \rowcolor[HTML]{FFFFFF}
        Developers & $>=$ 7 \\
        \rowcolor[HTML]{F3F3F3} 
        Pull Requests & $>$ 0 \\
        \rowcolor[HTML]{FFFFFF}
        Commits & $>$ 20 \\
        \rowcolor[HTML]{F3F3F3} 
        Releases & $>$ 1 \\
        \rowcolor[HTML]{FFFFFF}
        Issues & $>$ 10 \\
        \rowcolor[HTML]{F3F3F3} 
        Duration & $>$ 1 year \\
        \rowcolor[HTML]{FFFFFF}
        Has Travis CI & True \\
        \rowcolor[HTML]{F3F3F3}
        Total Builds & $>=$ 500 \\
        \rowcolor[HTML]{FFFFFF}
        Useful Builds & $>=$ 100 \\
        \rowcolor[HTML]{F3F3F3} 
        Failed Test Cases & $>=$ 50 
    \end{tabular}
\end{wraptable}

For our study, we implement the above algorithms to discover the best approach for the closed-source project and the open-source projects.

\section{Methods} \label{Methods}
Our overall experimental framework is described in~\fig{System Framework}. This section offers details on that framework.

\subsection{Data Collection} \label{DataCollection}

\begin{table*}[t!]
    \centering
    \caption{Summary of projects used in this study. (IQR = (75-25)th percentile).}
    \label{tbl:projectSummary}
    \scriptsize
    \centering
     \begin{tabular}{cc}
        \begin{tabular}{l|r|r|r|r}
           \rowcolor{gray!25}
            {\textbf{Feature}} & {\textbf{Min}} & {\textbf{Median}} & {\textbf{IQR}} & {\textbf{Max}} \\
            \hline
            \rowcolor[HTML]{FFFFFF}
            Developers & 8 & 39 & 57 & 188 \\ 
            \rowcolor[HTML]{F3F3F3}
            Commits & 2658 & 6189 & 10067 & 43627 \\
            \rowcolor[HTML]{FFFFFF}
            Releases & 1 & 15 & 21 & 167 \\
            \rowcolor[HTML]{F3F3F3}
            Issues & 310 & 827 & 667 & 3047 \\
            \rowcolor[HTML]{FFFFFF}
            Duration (week) & 137 & 292 & 197 & 529\\
            \rowcolor[HTML]{F3F3F3}
            Total Builds & 758 & 5094 & 5017 & 24692\\
            \rowcolor[HTML]{FFFFFF}
            Useful Builds & 193 & 719 & 991 & 7579\\
            \rowcolor[HTML]{F3F3F3}
            Failed Test Cases & 74 & 530 & 261 & 3554\\
        \end{tabular}
       &
        \begin{tabular}{l|r|r|r|r}
              \rowcolor{gray!25}
            {\textbf{Feature}} & {\textbf{Min}} & {\textbf{Median}} & {\textbf{IQR}} & {\textbf{Max}} \\
            \hline
            \rowcolor[HTML]{FFFFFF}
            Developers & 24 & 124 & 258 & 4020 \\
            \rowcolor[HTML]{F3F3F3}
            Commits & 969 & 14446 & 20939 & 77152 \\
            \rowcolor[HTML]{FFFFFF}
            Releases & 23 & 95 & 179 & 426 \\
            \rowcolor[HTML]{F3F3F3}
            Issues & 192 & 2369 & 2882 & 13848 \\
            \rowcolor[HTML]{FFFFFF}
            Duration (week) & 342 & 470 & 81 & 636\\
            \rowcolor[HTML]{F3F3F3}
            Total Builds & 206 & 2703 & 2597 & 19447\\
            \rowcolor[HTML]{FFFFFF}
            Useful Builds & 117 & 262 & 406 & 8794\\
            \rowcolor[HTML]{F3F3F3}
            Failed Test Cases & 50 & 93 & 111 & 5517\\
        \end{tabular}
        \\ \\
        \centering(a) Summary of 10 CS projects & \centering(b) Summary of 20 SE projects
    \end{tabular}
\end{table*}

For \revised{the} closed-source project, we use the data set from Yu et al.~\cite{yu2019terminator}. For \revised{the} open-source projects, we search \revised{on} GitHub.
Many projects in GitHub are very small or are out of maintenance, which may not have enough information for our experiments. To avoid these traps, we implement the GitHub ``sanity check'', which is introduced in the literature~\cite{kalliamvakou2014promises, munaiah2017curating,tu2020changing}. Our selection criteria is shown in \tbl{SanityCheck}.
Most of \revised{the} conditions in~\tbl{SanityCheck} are straight forward, but the last four conditions need explanation:
\bi
    \item
    \textbf{Has Travis CI:} We use the Travis CI API for collecting \revised{the} repository and build log information. Travis CI can let project developers test their applications and record testing information. Therefore, our ideal projects must implement Travis CI for the testing purpose.
    \item
    \textbf{Total Builds:} We use Travis CI to collect the execution history of test cases in each build. Travis will automatically trigger test suites after each build. However, for some builds that software developers skip the Travis manually, we discard these builds because they do not trigger the test suits. The rest builds are called {\em total builds}. This sanity check \revised{criterion} helps us to avoid small projects. 
    \item
    \textbf{Useful Builds:} \revised{In the} total builds, there are three types of builds. \revised{The first} is the passing builds. Since \revised{our target is} to prioritize failed test cases, we ignore these builds in this study because they have no failed test cases. Ignoring these builds will not change the orders of test cases. \revised{The second} is the broken builds with no \revised{failed test cases}. We discard these broken builds since \revised{the failure is not caused by the failed test cases}. \revised{The third} is the broken builds with tests failed. These builds are called ``failing builds'', and are marked as useful builds in this study. This sanity check \revised{criterion} helps us to avoid projects that do not have enough information for this study.
    \item
    \textbf{Failed Test Cases:} We count all failed test cases in the entire project. If a project has a very small number of failed test cases, then such a project is not suitable for our experiments.
\ei
\revised{To} ensure a diversity of \revised{the} open-source projects, we \revised{divide} the projects found in this way into different populations:
\bi
\item
We \revised{explore} the ``usual suspects''; i.e. projects that satisfy the sanity check \revised{criterion}
\revised{in} \tbl{SanityCheck}. Note that many of these projects have been used before in other publications.
We call this group \revised{as} the general software engineering group (hereafter, SE).
\item
We also \revised{explore} software from the computational science community. Computational Science (hereafter, CS) field studies and develops software to explore astronomy, astrophysics, chemistry, economics, genomics, molecular biology, oceanography, physics, political science, and many engineering fields~\cite{tu2020changing}.  
\ei
After the above analysis, we find ten projects from computational science and twenty projects from software engineering that suitable for our analysis: see \tbl{projectSummary}.

As to our closed-source case study data, this is the same data used in  Yu et al. case study~\cite{yu2019terminator}.
For reasons of corporate confidentiality,   
detailed  information is not publicly available. However, we can make some  general comments:
\bi
\item 
Our closed-source data comes from the nightly regression test suites which executed by LexisNexis.
This data is from dozens of projects, with hundreds of developers, working in multiple locations around the globe, all using the same cloud-based testing service.
\item
LexisNexis is a corporation providing computer-assisted legal research (CALR) as well as business research and risk management services~\cite{lnnyt1, lnnyt}. LexisNexis provides regulatory, legal, business information and analytic to the legal community. As of 2006, LexisNexis company had the world’s largest electronic database for legal and public-records related information~\cite{lnnyt2}.  
\item
The LexisNexis platform is maintained by a set of automated UI test suites. Those test suites simulate user behaviors on the interface of the platform and detect potential failures of the underlying micro-services whenever the system is
modified and rebuilt~\cite{yu2019terminator}.
\ei

\subsection{Data Preprocessing} \label{DataPreprocessing}
We \revised{use} the Travis CI API to extract GitHub repository information, (such as unique build id, commit id, and the starting time of the build), and build log information, (such as unique build id, build log status, and failed test cases). By matching the unique id in these two data sets, we can form our experimental data set. Since our target is prioritizing test cases in the new build by using previous execution history, the order of builds by time matters in this experiment. Fortunately, in most cases, Travis CI API will return test builds in consecutive order. Therefore, we only need to make small modifications to the data we collected. After we obtained information on failed test cases and test builds, we used a Python script to transfer the repository data and the build log data to the build-to-test tests record table for each project.

\begin{table*}[!t]
    \centering
    \caption{Information of Test Case Prioritization Algorithms}
    \label{tbl:TCPAlgorithm}
    \scriptsize
    \begin{tabular}[width=\textwidth]{crcl}
        \rowcolor{gray!25}
        {\textbf{Group ID}} & {\textbf{Information Utilized}} & {\textbf{Algorithm}} & {\textbf{Algorithm Description}} \\
        \hline
        \multirow{2}{*}{A} & \multirow{2}{*}{None} & A1 & Prioritize test cases randomly.\\ 
         & & A2 & Prioritize test cases optimally.\\
        \hline
        \multirow{4}{*}{B} &   \multirow{4}{*}{Execution History} & B1 & Prioritize test cases by the information of time since last failure.\\
         & & B2 & Prioritize test cases by the failure rate.\\
         & & B3 & Prioritize test cases by Exponential Decay Metrics. \\
         & & B4 & Prioritize test cases by ROCKET Metrics.\\
        \hline
        \multirow{2}{*}{C} & \multirow{2}{*}{Execution History, Feedback Information} & C1 & Prioritize test cases by co-failure information.\\
         & & C2 & Prioritize test cases by flipping history.\\
        \hline
        D & Execution History, Feedback Information & D1 & Prioritize test cases by TERMINATOR with execution history feature. 
    \end{tabular}
\end{table*}

\subsection{Performance Metric} \label{PerformanceMetric}
For the evaluation of prioritization algorithms, we implement fault detection rates. Rothermel et al.~\cite{rothermel2001prioritizing}  state that improved fault detection rate provides feedback faster than usual, which allows developers to correct faults earlier than normal time~\cite{rothermel2001prioritizing}. Their preferred measurement is called the weighted average of the percentage of faults detected (APFD). APFD calculates the area inside the curve that interpolates the gain in the percentage of detected faults~\cite{rothermel2001prioritizing}. It is calcuated
as follows:
\begin{equation}
    APFD = 1 - \frac{TC_1 + TC_2 + \cdots + TC_m}{nm} + \frac{1}{2n}
\end{equation}
where: 
\bi
    \item
    $TC_i$: The rank $i$ of the test case after prioritization that reveals fault.
    \item
    $m$: Total number of faults that are revealed in current test run.
    \item
    $n$: Total number of test cases in the current test run.
\ei
APFD ranges from 0 percent to 100 percent. A higher APFD value represents a larger area under the curve, which means higher fault detection rate, or better test case prioritization.

In APFD, all test cases are presumed to have the same execution time. Since the cost of test cases in GitHub projects is hard to be collected, APFD is the most suitable performance metric in our experiment.

\subsection{Test Case Prioritization Algorithms} \label{TCPAlgorithms}

Our study implements the nine prioritization algorithms found in the literature review of \S\ref{HowStudyTCP}.
While all these rely on execution history,  they prioritize test cases in different ways. We group these algorithms into Group A, B, C, and D according to the kinds of information that they use.
 
\bi
    \item
    \textbf{Group A}: Group A contains 2 approaches that prioritize test cases with no information gain. These two algorithms are baseline methods that are used for comparison.
    \item
    \textbf{Group B}: Group B includes 4 approaches that prioritize test cases only by their own execution history. They sort metrics to reorder the test cases before each test run.
    \item
    \textbf{Group C}: Group C has 2 approaches that prioritize test cases by correlations between two test cases. Two test cases have a large probability to have the same outcomes if they are highly correlated.
    \item
    \textbf{Group D}: Group D contains the proposed active learning based framework TERMINATOR~\cite{yu2019terminator}. TERMINATOR trains the SVM model with execution history when the first fault is detected.
\ei
\tbl{TCPAlgorithm} shows the detailed group division and a brief description of each algorithm. The \textbf{information utilized}   shows what history details \revised{are} used by each \revised{algorithm}.

Before we enter the detailed explanation of each algorithm we explored, there are some fundamental test case prioritization terms used in this study should be clarified. 
\bi
    \item In this study, a ``test run'' \revised{or a ``test build''} is defined as a single useful build in the projects. We start to record the APFD score for each test run after the fifth test run as Yu et al. did in their work~\cite{yu2019terminator}. This indicates that a project with $n$ useful builds will have $n-5$ recorded APFD scores. 
    \item For each test run, test case prioritization schemes will use all previous useful builds as the indicators to prioritize test cases.
\ei

In the rest of this section, we will explain how each algorithm \revised{prioritizes} test cases in each test build.
To clearly illustrate how each algorithm works,
we construct two small version of test case tables in Table~\ref{tbl:Example}, which have four test cases ($T_1$ - $T_4$) and five executed test builds ($B_1$ - $B_5$) in each example. Our target in these examples are prioritizing test cases for the sixth build. We put two examples here because (a) some algorithms may \revised{result in the} same orders in the small example. \revised{The} difference can be distributed by comparing two examples and (b) most of the test cases in example A have \revised{a} different number of failures in 5 builds, while most of the test cases in example B have \revised{a} similar number of failures, but failures are in \revised{the} different builds.

In these tables, \xmark \hspace{1mm}indicates failed testing \revised{results}, and \cmark \hspace{1mm}indicates passed testing \revised{results}. We assume all test cases have the same cost in \revised{these examples because of the real situation}.

\begin{table}[!ht]
    \centering
    \footnotesize
    \caption{Examples for Prioritization Algorithms}
    \label{tbl:Example}
    \begin{tabular}{c}
        \begin{tabular}{c|c c c c c|c}
        Test Case & $B_1$ & $B_2$ & $B_3$ & $B_4$ & $B_5$ & $B_6$ \\
        \hline  
        $T_1$ & \xmark & & & & & \cmark \\
        $T_2$ & \xmark & \xmark & & \xmark & & \xmark \\
        $T_3$ & & \xmark & \xmark & & & \cmark \\
        $T_4$ & & \xmark & & \xmark & \xmark & \xmark 
        \end{tabular} \\
        (a) Example A \\
        \begin{tabular}{c|c c c c c|c}
        Test Case & $B_1$ & $B_2$ & $B_3$ & $B_4$ & $B_5$ & $B_6$ \\
        \hline  
        $T_1$ & & \xmark & \xmark & & & \cmark \\
        $T_2$ & \xmark & & & \xmark & & \xmark \\
        $T_3$ & & & \xmark & & \xmark & \xmark \\
        $T_4$ & \xmark & \xmark & \xmark & & & \cmark 
        \end{tabular} \\
        (b) Example B
    \end{tabular}
\end{table}

    \textbf{A1:} This algorithm implements no information. It prioritizes test cases in random order. This is the baseline method since all prioritization algorithms should have better performance than A1.  
    
\textbf{A2:} This algorithm uses the historical record of failed tests
to sort the tests. 
In \tbl{Example} Example A, A2 will execute $T_2$ and $T_4$ randomly before $T_1$ and $T_3$ since they will reveal faults in the current test run. In Example B, A2 will execute $T_2$ and $T_3$ first, and then $T_1$ and $T_4$. We call A2 the {\em omniscient algorithm} since it uses information that is unavailable before prioritizing new tests. Note that if A1 represents the dumbest prioritization, then the omniscient A2 algorithm represents the best possible decisions. In the rest of this paper, we compare all results against A1 and A2 since that will let us \revised{compare the} prioritization \revised{results} against the \revised{worst} and \revised{the} most omniscient decisions.

\textbf{B1:} B1 uses the time since the last failure. A test case with less consecutive non-failure builds will be assigned with higher priority~\cite{elbaum2014techniques, hemmati2015prioritizing}. 
\revised{In the example A in \tbl{Example}}, $T_4$ has 0 consecutive non-failure build since it failed in $B_4$. Thus $T_4$ will be executed first. After that, $T_2$ has 1 consecutive failure, so it will be executed next. Moreover, $T_3$ and $T_1$ have 2 and 4 consecutive non-failure builds. Therefore, the final order in Example A is $\{T_4, T_2, T_3, T_1\}$. When \revised{the} same ranking scheme is applied \revised{to} Example 2, \revised{the prioritized order will become to} $\{T_3, T_2, T_1, T_4\}$. The metrics and the corresponding orders of \revised{the} two examples are shown in~\tbl{B1Example}.

\begin{table}[!ht]
    \centering
    \footnotesize
    \caption{Metrics and Orders of Algorithm B1}
    \label{tbl:B1Example}
    \begin{tabular}{c||c|c||c|c}
         & \multicolumn{2}{c||}{Example A} & \multicolumn{2}{c}{Example B} \\
         \hline
         & Metric & Order & Metric & Order \\
         \hline
         $T_1$ & 4 & 4 & 2 & 3 \\
         \hline
         $T_2$ & 1 & 2 & 1 & 2 \\
         \hline
         $T_3$ & 2 & 3 & 0 & 1 \\
         \hline
         $T_4$ & 0 & 1 & 2 & 4 \\
    \end{tabular}
\end{table}


\textbf{B2:} B2 uses the value of failure rate in metrics to prioritize test cases~\cite{fazlalizadeh2009prioritizing}. Failure rate is defined as: 
\begin{center}
(total number of failed builds) / (total test builds)
\end{center}
A test case with higher failure rate has \revised{been} executed earlier. 
In example A, we can calculate the failure rate of $T_1$ to $T_4$ with $1/5=0.2$, $3/5=0.6$, $2/5=0.4$, and $3/5=0.6$. Thus, the algorithm will result $\{T_2, T_4, T_3, T_1\}$. In example B, the failure rates of $T_1$ to $T_4$ are $2/5=0.4$, $2/5=0.4$, $2/5=0.4$, $3/5=0.6$. We can find 3 test cases have \revised{the} same failure rate. Thus, we will order them randomly. The metrics and the corresponding orders of \revised{the} two examples are shown in~\tbl{B2Example}.

\begin{table}[!ht]
    \centering
    \footnotesize
    \caption{Metrics and Orders of Algorithm B2}
    \label{tbl:B2Example}
    \begin{tabular}{c||c|c||c|c}
         & \multicolumn{2}{c||}{Example A} & \multicolumn{2}{c}{Example B} \\
         \hline
         & Metric & Order & Metric & Order \\
         \hline
         $T_1$ & 0.2 & 4 & 0.4 & 2 \\
         \hline
         $T_2$ & 0.6 & 1 & 0.4 & 4 \\
         \hline
         $T_3$ & 0.4 & 3 & 0.4 & 3 \\
         \hline
         $T_4$ & 0.6 & 2 & 0.6 & 1 \\
    \end{tabular}
\end{table}

\textbf{B3:} B3 implements the ``Exponential Decay Metric'' (mentioned earlier in this paper)
to calculate the ranking values of test cases~\cite{kim2002history, fazlalizadeh2009prioritizing}:
    \begin{align*}
        P_0 &= B_1 \\
        P_k &= \alpha B_k + (1-\alpha)P_{k-1}, 0 \leq \alpha \leq 1, k \geq 1
    \end{align*}
where variables in these equations are defined as:
\bi
    \item $B_i$: The test result in build $i$. $B_i = 0$ if test passed and $B_i = 1$ if test failed.
    \item $\alpha$: The learning rate. In our experiments, we test the value of $\alpha$ from 0 to 1 with 0.1 interval. We find $\alpha = 0.9$ reaches highest performance.
    \item $P_j$: Exponential Decay value of test case $j$.
\ei

A test case with higher Exponential Decay value will be executed earlier. In example A, the Exponential Decay values for $T_1$ to $T_4$ are $\{0.0001, 0.091, 0.0099, 0.9909\}$. 
In example B, the Exponential Decay values for $T_1$ to $T_4$ are $\{0.0099, 0.0901, 0.909, 0.01\}$. The metrics and the corresponding orders of \revised{the} two examples are shown in~\tbl{B3Example}.
\begin{table}[!ht]
    \centering
    \footnotesize
    \caption{Metrics and Orders of Algorithm B3}
    \label{tbl:B3Example}
    \begin{tabular}{c||c|c||c|c}
         & \multicolumn{2}{c||}{Example A} & \multicolumn{2}{c}{Example B} \\
         \hline
         & Metric & Order & Metric & Order \\
         \hline
         $T_1$ & 0.0001 & 4 & 0.0099 & 4 \\
         \hline
         $T_2$ & 0.091 & 2 & 0.0901 & 2 \\
         \hline
         $T_3$ & 0.0099 & 3 & 0.909 & 1 \\
         \hline
         $T_4$ & 0.9909 & 1 & 0.01 & 3 \\
    \end{tabular}
\end{table}

\textbf{B4:} B4 prioritizes test cases by implementing the ROCKET Metrics~\cite{marijan2013TCP, cho2016history}. In the ROCKET Metrics, prioritization value $P=\{P_1, P_2, \cdots, P_n\}$ is calculated as follow: 
    \begin{align*}
        w_{i} &= \left\{ \begin{array}{l}
                    0.7, \mbox{  if $i=1$}\\
                    0.2, \mbox{  if $i=2$}\\
                    0.1, \mbox{  if $i \geq 3$}
                \end{array} \right. \\
        P_i &= \sum_{j=1}^{i-1} B_j * w_{i-j}
    \end{align*}
where variables in this system are defined as:
\bi
    \item $B_i$: The test result in build $i$. $B_i = 0$ if test passed and $B_i = 1$ if test failed.
    \item $P_j$: The ROCKET value of test case $j$.
\ei
The prioritization value $P$ will be ranked in descending order. Test cases will be executed in the ranked orders. In example A in~\tbl{Example}, this scheme will calculate the results of $T_1$ to $T_4$ \revised{as} $\{0.1, 0.4, 0.2, 1.0\}$, and in example B, ROCKET Metric of $T_1$ to $T_4$ is $\{0.2, 0.3, 0.8, 0.3\}$. We can observe that the test case \revised{failed} in the previous two builds or the test case with more failures in the past will have \revised{a} higher \revised{score}. The metrics and the corresponding orders \revised{of the} two examples are shown in~\tbl{B4Example}.
\begin{table}[!ht]
    \centering
    \footnotesize
    \caption{Metrics and Orders of Algorithm B4}
    \label{tbl:B4Example}
    \begin{tabular}{c||c|c||c|c}
         & \multicolumn{2}{c||}{Example A} & \multicolumn{2}{c}{Example B} \\
         \hline
         & Metric & Order & Metric & Order \\
         \hline
         $T_1$ & 0.1 & 4 & 0.2 & 4 \\
         \hline
         $T_2$ & 0.4 & 2 & 0.3 & 3 \\
         \hline
         $T_3$ & 0.2 & 3 & 0.8 & 1 \\
         \hline
         $T_4$ & 1.0 & 1 & 0.3 & 2 \\
    \end{tabular}
\end{table}
 
\textbf{C1}: The C1 algorithm was introduced by Zhu et al. in 2018. They consider the past test co-failure distributions in test case prioritization~\cite{zhu2018test}. Making two test cases as a pair of tests, the co-failure score is calculated by:
    \begin{align*}
        Score(t) = prevScore(t) + (P(t = fail|t_{finished}) - 0.5)
    \end{align*}
where variables in this equation are defined as:
\bi
    \item
    $t$: Test cases that are not executed.
    \item
    $t_{finished}$: Test case that executed just now.
    \item
    $Score(t)$: Score of test case $t$ in current test run.
    \item
    $prevScore(t)$: Score of test case $t$ in previous test run.
\ei
A higher score in this approach means highly correlated with the executed test cases. 

The test case with \revised{the} higher score in this approach means highly correlated with the executed test cases. In example A in~\tbl{Example}, by given $T_2$ failed initially, $T_1$ failed one time when $T_2$ failed in the history. Thus, the score of $T_1$ is updated to $0 + 1/3 - 0.5 = -0.17$. By using the same way to calculate $T_3$ and $T_4$, the scores of $T_1$, $T_3$, and $T_4$ will be updated from initial 0 to $\{-0.17, -0.17, 0.17\}$. Since $T_4$ has the highest score, $T_4$ is highly correlated with $T_2$ in this example. Therefore, $T_4$ will be executed next. After $T_4$ being executed and failed, the scores of $T_1$ and $T_3$ is updated to $\{-0.67, -0.34\}$. Since $T_3$ has \revised{a} higher score than $T_1$, this scheme will order $T_3$ before $T_1$. In example B, given $T_3$ failed, the scores of \revised{the} rest test cases \revised{are} updated to $\{0, -0.5, 0\}$. Since $T_1$ and $T_4$ have same score, we will randomly select $T_4$. After $T_4$ is being executed and passed, the score of $T_1$ is updated from 0 to -0.5 since $P(t=fail|t_{finished}=pass) = 0$, and the score of $T_2$ is updated from -0.5 to -0.5 since $P(t=fail|t_{finished}=pass) = 0.5$. Therefore, $T_2$ will be executed randomly before $T_1$ because they have the same score. The detailed scores and corresponding orders of \revised{the} two examples are shown in~\tbl{C1Example}.

\begin{table}[!ht]
    \centering
    \footnotesize
    \caption{Metrics and Orders of Algorithm C1}
    \label{tbl:C1Example}
    \begin{tabular}{c||c|c||c|c}
         & \multicolumn{2}{c||}{Example A} & \multicolumn{2}{c}{Example B} \\
         \hline
         & Metric & Order & Metric & Order \\
         \hline
         $T_1$ & -0.17,-0.67 & 4 & 0,-0.5 & 4 \\
         \hline
         $T_2$ & -,- & 1 & -0.5,-0.5 & 3 \\
         \hline
         $T_3$ & -0.17,-0.34 & 3 & -,- & 1 \\
         \hline
         $T_4$ & 0.17,- & 2 & 0,- & 2 \\
    \end{tabular}
\end{table}

\textbf{C2}: C2 algorithm is proposed by Cho et al. in 2016. They define that two test cases are highly correlated if their testing results change to the opposite status (flip) in two consecutive test runs~\cite{cho2016history}. Moreover, they utilize the ROCKET method to find the first failed test case. 

In example A in~\tbl{Example}, the ROCKET approach \revised{locates} $T_4$ first. After $T_4$ executed and failed, we can find $T_1$ flips one time \revised{by comparing} to $T_4$ in $B_1$ to $B_2$ because $T_1$ and $T_4$ both change their results to the opposite \revised{side}. By using the same way to calculate, the flipping history for $T_1$ to $T_3$ is $\{1, 2, 2\}$. Thus, $T_2$ and $T_3$ will be randomly selected. Assume $T_2$ is assigned to second order, $T_1$ and $T_3$ will update their flipping history to $\{0, 1\}$. Hence $T_3$ will be executed before $T_1$. In example B, \revised{the} ROCKET approach will assign $T_3$ to the first order. The flipping history of $T_1$, $T_2$, and $T_4$ will update to $\{1, 2, 1\}$. After $T_2$ executed and failed, the flipping history will update to $\{2, 1 \}$. Thus $T_1$ will be executed before $T_4$. The detailed scores and corresponding orders of \revised{the} two examples are shown in~\tbl{C2Example}.

\begin{table}[!ht]
    \centering
    \footnotesize
    \caption{Metrics and Orders of Algorithm C2}
    \label{tbl:C2Example}
    \begin{tabular}{c||c|c||c|c}
         & \multicolumn{2}{c||}{Example A} & \multicolumn{2}{c}{Example B} \\
         \hline
         & Metric & Order & Metric & Order \\
         \hline
         $T_1$ & 1,0 & 4 & 1,2 & 3 \\
         \hline
         $T_2$ & 2,- & 2 & 2,- & 2 \\
         \hline
         $T_3$ & 2,1 & 3 & -,- & 1 \\
         \hline
         $T_4$ & -,- & 1 & 1,1 & 4 \\
    \end{tabular}
\end{table}

\textbf{D1}: The last algorithm TERMINATOR in our experiments was proposed  by Yu et al. in 2019. TERMINATOR implements an active learning based framework~\cite{yu2019terminator}. This approach uses execution history as features to incrementally train a support vector machine classifier. Uncertainty sampling\footnote{Execute the test case with the most uncertain predicted probability.} is applied until the number of detected faults exceeds some threshold $N_1$. After that, certainty sampling\footnote{Execute the test case with the highest predicted probability.} is utilized until all test cases are prioritized. 

In the examples, we set the threshold $N_1$ to 2. In example A in~\tbl{Example}, with $T_2$ being randomly executed, and labeled as \revised{a} failed test case, the algorithm randomly presume $T_1$ as a non-relevant test case. After that, an SVM learner is trained by using the execution history of $T_1$ and $T_2$, with $T_2$ labeled as failed and $T_1$ labeled as passed. Here, we assume that the fitting results of $T_1$, $T_3$, and $T_4$ are $\{0.3, 0.6, 0.55\}$. Since the algorithm does not hit the threshold, uncertainty sampling is applied to this result. In this case, $T_4$ is selected because it is the most uncertain sample (closest to 0.5). After that, with more evidence, we assume $T_1$ and $T_3$ have prediction result $\{0.2, 0.8\}$. Since the number of failed test cases exceeds the threshold, certainty sampling will be applied. $T_3$ will be executed next because it has the highest predicted probability to reveal the fault. The final order in this example is $\{T_2, T_4, T_3, T_1\}$. In example B, if $T_4$ is randomly selected first and passed, the algorithm will randomly select another test case $T_3$. Since $T_3$ failed, the algorithm will start learning as example A does. The metrics and the corresponding orders \revised{of the} two examples are shown in~\tbl{D1Example}.

\begin{table}[!ht]
    \centering
    \footnotesize
    \caption{Metrics and Orders of Algorithm D1}
    \label{tbl:D1Example}
    \begin{tabular}{c||c|c||c|c}
         & \multicolumn{2}{c||}{Example A} & \multicolumn{2}{c}{Example B} \\
         \hline
         & Metric & Order & Metric & Order \\
         \hline
         $T_1$ & 0.3,0.2 & 4 & 0.75,0.3 & 4 \\
         \hline
         $T_2$ & -,- & 1 & 0.6,- & 3 \\
         \hline
         $T_3$ & 0.6,0.8 & 3 & -,- & 2 \\
         \hline
         $T_4$ & 0.55,- & 2 & -,- & 1 \\
    \end{tabular}
\end{table}

\subsection{Statistical Methods} \label{StatisticalMethod}
In our study, we report the median and interquartile range (which show 50th percentile and 75th-25th percentile), of APFD results for entire test runs. We collect median and interquartile range values for each of the projects. 

To make comparisons among all algorithms on a single project, we implement the Scott-Knott analysis~\cite{mittas2012ranking}. In summary,
\revised{by} using Scott-Knott, algorithms are sorted by their performance.
\revised{After that, they are} assigned \revised{to} different ranks
if \revised{the performance of the algorithm at position $i$} is significantly 
different to the algorithm at position $i-1$.

To be more precise, Scott-Knott sorts the list of treatments (in this paper, the prioritization algorithms) by their median score. After the sorting, it then splits the list into two sub-lists. The goal for such a split is to maximize the expected value of differences in the observed performances before and after division~\cite{xia2018hyperparameter}. For example, in our study, we implement 9 prioritization approaches in list $l$ as treatments, then the possible divisions of $l_1$ and $l_2$ are $(l_1, l_2) \in \{(1,8), (2,7), (3,6), (4,5), (5,4), (6,3), (7,2), (8,1)\}$. Scott-Knott analysis then declares one of the above divisions to be the best split. The best split should maximize the difference $E(\Delta)$ in the expected mean value before and after the split:
\begin{equation}
    E(\Delta) = \frac{|l_1|}{|l|}abs(\overline{l_1} - \overline{l})^2 + \frac{|l_2|}{|l|}abs(\overline{l_2} - \overline{l})^2
\end{equation}
where:
\bi
    \item
    $|l|$, $|l_1|$, and $|l_2|$: Size of list $l$, $l_1$, and $l_2$.
    \item
    $\overline{l}$, $\overline{l_1}$, and $\overline{l_2}$: Mean value of list $l$, $l_1$, and $l_2$.
\ei
After the best split is declared by the formula above, Scott-Knott then implements some statistical hypothesis tests to check whether the division is useful or not. Here ``useful'' means $l_1$ and $l_2$ differ significantly by applying hypothesis test $H$. If the division is checked as a useful split, the Scott-Knott analysis will then run recursively on each half of the best split until no division can be made. In our study, hypothesis test $H$ is the cliff's delta non-parametric effect size measure. Cliff's delta quantifies the number of difference between two lists of observations beyond p-values interpolation~\cite{macbeth2011cliff}. The division passes the hypothesis test if it is not a ``small'' effect ($Delta \geq 0.147$). The cliff's delta non-parametric effect size test explores two lists $A$ and $B$ with size $|A|$ and $|B|$:
\begin{equation}
    Delta = \frac{\sum\limits_{x \in A} \sum\limits_{y \in B} \left\{ \begin{array}{l}
                    +1, \mbox{   if $x > y$}\\
                    -1, \mbox{   if $x < y$}\\
                    0,  \mbox{   if $x = y$}
                \end{array} \right.}{|A||B|}
\end{equation}
In this expression, cliff's delta estimates the probability that a value in list $A$ is greater than a value in list $B$, minus the reverse probability~\cite{macbeth2011cliff}. This hypothesis test and its effect size is supported by Hess and Kromery~\cite{hess2004robust}.

\begin{table}[!t]
\caption{Scott-Knott analysis \revised{of our} proprietary data. In this table, ``med'' denotes median;    \colorbox{blue!10}{blue row} shows   D1 algorithm, while \colorbox{red!10}{red row} shows     B1/B3.}
\label{tbl:SK_LN}
\centering
\scriptsize
\begin{adjustbox}{max width=0.48\textwidth}
\begin{tabular}{ccccl}
    \rowcolor[gray]{1} \textbf{rank} & \textbf{what} & \textbf{med} & \textbf{IQR} & \\
    \hline
    1 & A1 & 0.50 & 0.02 & \quart{49}{2}{50}{100}\\
    2 & C2 & 0.69 & 0.06 & \quart{65}{6}{69}{100}\\
    \rowcolor{red!10}\textbf{3} & \textbf{B1} & \textbf{0.70} & \textbf{0.08} & \quart{66}{8}{70}{100}\\
    \rowcolor{red!10}\textbf{3} & \textbf{B3} & \textbf{0.72} & \textbf{0.08} & \quart{68}{8}{72}{100}\\
    4 & B2 & 0.74 & 0.08 & \quart{71}{8}{74}{100}\\
    4 & B4 & 0.75 & 0.08 & \quart{70}{8}{75}{100}\\
    5 & C1 & 0.79 & 0.09 & \quart{73}{9}{79}{100}\\
    \rowcolor{blue!10}\textbf{5} & \textbf{D1} & \textbf{0.80} & \textbf{0.14} & \quart{71}{14}{80}{100}\\
    6 & A2 & 0.96 & 0.08 & \quart{89}{8}{96}{100}
\end{tabular}
\end{adjustbox}
\end{table}

\section{Results} \label{Results}

In this section, we will show our experimental results and answer RQs with these results.
Note that RQ1 only shows results for the same closed-source project studied in the TERMINATOR paper~\cite{yu2019terminator} while
RQ2 states the experimental results for 30 open-source projects.

\subsection{What is the best algorithm for closed-source project? (RQ1)} \label{RQ1}

To answer RQ1, we reproduce the Yu et al. study by implementing the prioritization approaches we find from the previous literature in their data set~\cite{yu2019terminator}. Note that, for this data, Yu et al. recommended TERMINATOR (which we call the D1 prioritization algorithm).

\tbl{SK_LN} shows our simulation results of 9 prioritization algorithms. We record APFD result of each test run, and calculate median value and interquartile range of APFD for all test runs. An algorithm with \revised{a} higher APFD value has \revised{a} better performance. 
As described in \S\ref{StatisticalMethod}, \revised{two} algorithms differ significantly if they separate in different ranks \revised{in} the Scott-Knott analysis.

\begin{table*}[!b]
    \caption{Scott-Knott analysis results for 10 open-source computational science projects. In these tables \colorbox{blue!10}{blue row} denotes the performance of D1 algorithm, while \colorbox{red!10}{red row} denotes the performance of B1/B3 approach.
    Note that B1/B3 is ranked the same as the omniscient A2 method \revised{in 8/10 cases}: see figures b,c,e,f,g,h,i,j.}
    \label{tbl:opensourceResultCS}
  \begin{center}
    \begin{adjustbox}{max width = .85\textwidth}

    \begin{tabular}{c@{}c@{}c}
        \begin{tabular}{ccccl}
            \cellcolor[gray]{1}\textbf{rank} & \cellcolor[gray]{1}\textbf{what} &     \cellcolor[gray]{1}\textbf{med} & \cellcolor[gray]{1}\textbf{IQR} & \cellcolor[gray]{1}\\
            \hline
            \rowcolor{white}  1 & A1 & 0.51 & 0.45 & \quart{25}{45}{51}{100} \\
            \rowcolor{blue!10} \textbf{1} & \textbf{D1} & \textbf{0.54} & \textbf{0.38} & \quart{31}{38}{54}{100} \\
            \rowcolor{white}  2 & B2 & 0.79 & 0.46 & \quart{47}{46}{79}{100} \\
            \rowcolor{white}  2 & C1 & 0.81 & 0.48 & \quart{45}{48}{81}{100} \\
            \rowcolor{white}  2 & B4 & 0.88 & 0.48 & \quart{49}{48}{88}{100} \\
            \rowcolor{white}  2 & C2 & 0.89 & 0.39 & \quart{57}{39}{89}{100} \\
            \rowcolor{red!10}   \textbf{3} & \textbf{B1} & \textbf{0.97} & \textbf{0.35} & \quart{64}{35}{97}{100} \\
            \rowcolor{red!10}   \textbf{3} & \textbf{B3} & \textbf{0.97} & \textbf{0.32} & \quart{67}{32}{97}{100} \\
            \rowcolor{white} 4 & A2 & 0.99 & 0.00 & \quart{99}{0}{99}{100} \\
        \end{tabular} & 
        \begin{tabular}{|ccccl}
            \cellcolor[gray]{1}\textbf{rank} & \cellcolor[gray]{1}\textbf{what} &     \cellcolor[gray]{1}\textbf{med} & \cellcolor[gray]{1}\textbf{IQR} & \cellcolor[gray]{1}\\
            \hline
            \rowcolor{white}1 &      A1 &    0.51 &  0.26 & \quart{41}{26}{51}{100} \\
            \rowcolor{blue!10}\textbf{1} &      \textbf{D1} &    \textbf{0.60} &  \textbf{0.36} & \quart{38}{36}{60}{100} \\
            \rowcolor{white}2 &      B2 &    0.81 &  0.34 & \quart{65}{34}{81}{100} \\
            \rowcolor{white}2 &      B4 &    0.82 &  0.34 & \quart{65}{34}{82}{100} \\
            \rowcolor{white}2 &      C2 &    0.84 &  0.40 & \quart{59}{40}{84}{100} \\
            \rowcolor{white}3 &      C1 &    0.98 &  0.20 & \quart{78}{20}{98}{100} \\
            \rowcolor{red!10}\textbf{4} &      \textbf{B1} &    \textbf{0.99} &  \textbf{0.20} & \quart{79}{20}{99}{100} \\
            \rowcolor{red!10}\textbf{4} &      \textbf{B3} &    \textbf{0.99} &  \textbf{0.19} & \quart{80}{19}{99}{100} \\
            \rowcolor{white}4 &      A2 &    0.99 &  0.15 & \quart{84}{15}{99}{100} \\
        \end{tabular} &
        \begin{tabular}{|ccccl}
            \cellcolor[gray]{1}\textbf{rank} & \cellcolor[gray]{1}\textbf{what} &     \cellcolor[gray]{1}\textbf{med} & \cellcolor[gray]{1}\textbf{IQR} & \cellcolor[gray]{1} \\
            \hline
            \rowcolor{white}1 &      A1 &    0.50 &  0.28 & \quart{34}{28}{50}{100} \\
            \rowcolor{blue!10} \textbf{2} &      \textbf{D1} &    \textbf{0.66} &  \textbf{0.48} & \quart{37}{48}{66}{100} \\
            \rowcolor{white}3 &      B2 &    0.92 &  0.33 & \quart{67}{33}{92}{100} \\
            \rowcolor{white}3 &      C2 &    0.94 &  0.28 & \quart{71}{28}{94}{100} \\
            \rowcolor{white}3 &      B4 &    0.95 &  0.27 & \quart{73}{27}{95}{100} \\
            \rowcolor{white}3 &      C1 &    0.96 &  0.24 & \quart{76}{24}{96}{100} \\
            \rowcolor{red!10}\textbf{4} &      \textbf{B1} &    \textbf{0.99} &  \textbf{0.16} & \quart{84}{16}{99}{100} \\
            \rowcolor{red!10}\textbf{4} &      \textbf{B3} &    \textbf{0.99} &  \textbf{0.16} & \quart{84}{16}{99}{100} \\
            \rowcolor{white}4 &      A2 &    1.00 &  0.01 & \quart{99}{1}{100}{100} \\
        \end{tabular} \\
        (a). Project Name: parsl/parsl & (b). Project Name: radical-sybertools/radical & (c). Project Name: yt-project/yt
        \\
        \\
        \begin{tabular}{ccccl}
             \cellcolor[gray]{1}\textbf{rank} & \cellcolor[gray]{1}\textbf{what} &     \cellcolor[gray]{1}\textbf{med} & \cellcolor[gray]{1}\textbf{IQR} & \cellcolor[gray]{1}
            \\
            \hline
            \rowcolor{white}1 &      A1 &    0.50 &  0.14 & \quart{42}{14}{50}{100} \\
            \rowcolor{white}2 &      B2 &    0.80 &  0.44 & \quart{51}{44}{80}{100} \\
            \rowcolor{blue!10}\textbf{2} &      \textbf{D1} &    \textbf{0.83} &  \textbf{0.34} & \quart{56}{34}{83}{100} \\
            \rowcolor{white}2 &      C2 &    0.83 &  0.32 & \quart{64}{32}{83}{100} \\
            \rowcolor{white}2 &      B4 &    0.85 &  0.37 & \quart{59}{37}{85}{100} \\
            \rowcolor{white}2 &      C1 &    0.89 &  0.39 & \quart{57}{39}{89}{100} \\
            \rowcolor{red!10}\textbf{3} &      \textbf{B1} &    \textbf{0.98} &  \textbf{0.16} & \quart{83}{16}{98}{100} \\
            \rowcolor{red!10}\textbf{3} &      \textbf{B3} &    \textbf{0.98} &  \textbf{0.15} & \quart{84}{15}{98}{100} \\
            \rowcolor{white}4 &      A2 &    0.99 &  0.01 & \quart{99}{1}{99}{100} \\
        \end{tabular} &
        \begin{tabular}{|ccccl}
            \cellcolor[gray]{1}\textbf{rank} & \cellcolor[gray]{1}\textbf{what} &     \cellcolor[gray]{1}\textbf{med} & \cellcolor[gray]{1}\textbf{IQR} & \cellcolor[gray]{1}\\
            \hline
            \rowcolor{white}1 &      A1 &    0.50 &  0.20 & \quart{40}{20}{50}{100} \\
            \rowcolor{blue!10}\textbf{2} &      \textbf{D1} &    \textbf{0.79} &  \textbf{0.40} & \quart{50}{40}{79}{100} \\
            \rowcolor{white}3 &      C2 &    0.95 &  0.13 & \quart{86}{13}{95}{100} \\
            \rowcolor{white}3 &      B2 &    0.95 &  0.13 & \quart{86}{13}{95}{100} \\
            \rowcolor{white}3 &      B4 &    0.96 &  0.10 & \quart{89}{10}{96}{100} \\
            \rowcolor{red!10}\textbf{4} &      \textbf{B1} &    \textbf{0.99} &  \textbf{0.02} & \quart{98}{2}{99}{100} \\
            \rowcolor{red!10}\textbf{4} &      \textbf{B3} &    \textbf{0.99} &  \textbf{0.01} & \quart{99}{1}{99}{100} \\
            \rowcolor{white}4 &      A2 &    1.00 &  0.01 & \quart{99}{1}{100}{100} \\
            \rowcolor{white} &          &         &       & \\
        \end{tabular} & 
        \begin{tabular}{|ccccl}
            \cellcolor[gray]{1}\textbf{rank} & \cellcolor[gray]{1}\textbf{what} &     \cellcolor[gray]{1}\textbf{med} & \cellcolor[gray]{1}\textbf{IQR} & \cellcolor[gray]{1}\\
            \hline
            \rowcolor{white}1 &      A1 &    0.50 &  0.41 & \quart{30}{41}{50}{100} \\
            \rowcolor{blue!10}\textbf{1} &      \textbf{D1} &    \textbf{0.53} &  \textbf{0.48} & \quart{30}{48}{53}{100} \\
            \rowcolor{white}2 &      C2 &    0.99 &  0.14 & \quart{86}{14}{99}{100} \\
            \rowcolor{white}2 &      B2 &    1.00 &  0.08 & \quart{92}{8}{100}{100} \\
            \rowcolor{white}2 &      B4 &    1.00 &  0.05 & \quart{95}{5}{100}{100} \\
            \rowcolor{red!10}\textbf{2} &      \textbf{B1} &    \textbf{1.00} &  \textbf{0.01} & \quart{100}{1}{100}{100} \\
            \rowcolor{red!10}\textbf{2} &      \textbf{B3} &    \textbf{1.00} &  \textbf{0.01} & \quart{100}{1}{100}{100} \\
            \rowcolor{white}2 &      A2 &    1.00 &  0.00 & \quart{100}{0}{100}{100} \\
            \rowcolor{white} &          &         &       & \\
        \end{tabular} \\
        (d). Project Name: mdanalysis/mdanalysis & (e). Project Name: unidata/metpy & (f). Project Name: materialsproject/pymatgen
        \\
        \\
        \begin{tabular}{ccccl}
            \cellcolor[gray]{1}\textbf{rank} & \cellcolor[gray]{1}\textbf{what} &     \cellcolor[gray]{1}\textbf{med} & \cellcolor[gray]{1}\textbf{IQR} & \cellcolor[gray]{1}\\
            \hline
            \rowcolor{white}1 &      A1 &    0.50 &  0.24 & \quart{38}{24}{50}{100} \\
            \rowcolor{blue!10}\textbf{2} &      \textbf{D1} &    \textbf{0.74} &  \textbf{0.39} & \quart{50}{39}{74}{100} \\
            \rowcolor{white}3 &      B2 &    0.91 &  0.24 & \quart{74}{24}{91}{100} \\
            \rowcolor{white}3 &      B4 &    0.92 &  0.23 & \quart{75}{23}{92}{100} \\
            \rowcolor{white}3 &      C2 &    0.93 &  0.20 & \quart{78}{20}{93}{100} \\
            \rowcolor{red!10} \textbf{4} &     \textbf{B1} &    \textbf{0.99} &  \textbf{0.12} & \quart{88}{12}{99}{100} \\
            \rowcolor{red!10}  \textbf{4} &     \textbf{B3} &    \textbf{0.99} &  \textbf{0.11} & \quart{89}{11}{99}{100} \\
            \rowcolor{white}4 &      A2 &    1.00 &  0.01 & \quart{99}{1}{100}{100} \\
        \end{tabular} & 
        \begin{tabular}{|ccccl}
            \cellcolor[gray]{1}\textbf{rank} & \cellcolor[gray]{1}\textbf{what} &     \cellcolor[gray]{1}\textbf{med} & \cellcolor[gray]{1}\textbf{IQR} & \cellcolor[gray]{1}\\
            \hline
            \rowcolor{white}1 &      A1 &    0.50 &  0.14 & \quart{42}{14}{50}{100} \\
            \rowcolor{blue!10}2 &      \textbf{D1} &    \textbf{0.83} &  \textbf{0.38} & \quart{55}{38}{83}{100} \\
            \rowcolor{white}2 &      B2 &    0.84 &  0.38 & \quart{60}{38}{84}{100} \\
            \rowcolor{white}3 &      B4 &    0.92 &  0.30 & \quart{69}{30}{92}{100} \\
            \rowcolor{white}3 &      C2 &    0.94 &  0.24 & \quart{75}{24}{94}{100} \\
            \rowcolor{red!10}\textbf{4} &      \textbf{B1} &    \textbf{1.00} &  \textbf{0.10} & \quart{90}{10}{100}{100} \\
            \rowcolor{red!10}\textbf{4} &      \textbf{B3} &    \textbf{1.00} &  \textbf{0.11} & \quart{89}{11}{100}{100} \\
            \rowcolor{white}4 &      A2 &    1.00 &  0.00 & \quart{100}{0}{100}{100} \\
        \end{tabular} &
        \begin{tabular}{|ccccl}
            \cellcolor[gray]{1}\textbf{rank} & \cellcolor[gray]{1}\textbf{what} &     \cellcolor[gray]{1}\textbf{med} & \cellcolor[gray]{1}\textbf{IQR} & \cellcolor[gray]{1}\\
            \hline
            \rowcolor{white}1 &      A1 &    0.50 &  0.29 & \quart{35}{29}{50}{100} \\
            \rowcolor{blue!10}\textbf{2} &      \textbf{D1} &    \textbf{0.69} &  \textbf{0.38} & \quart{47}{38}{69}{100} \\
            \rowcolor{white}3 &      C2 &    0.99 &  0.06 & \quart{94}{6}{99}{100} \\
            \rowcolor{white}3 &      B4 &    0.99 &  0.03 & \quart{97}{3}{99}{100} \\
            \rowcolor{white}3 &      B2 &    0.99 &  0.03 & \quart{97}{3}{99}{100} \\
            \rowcolor{red!10}\textbf{3} &      \textbf{B1} &    \textbf{1.00} &  \textbf{0.01} & \quart{99}{1}{100}{100} \\
            \rowcolor{red!10}\textbf{3} &      \textbf{B3} &    \textbf{1.00} &  \textbf{0.01} & \quart{99}{1}{100}{100} \\
            \rowcolor{white}3 &      A2 &    1.00 &  0.00 & \quart{100}{0}{100}{100} \\
        \end{tabular} \\
        (g). Project Name: reactionMechanism../RMG-Py & (h). Project Name: openforcefield/openforcefield & (i). Project Name: spotify/luigi
        \\
        \\
        &
        \begin{tabular}{ccccl}
            \cellcolor[gray]{1}\textbf{rank} & \cellcolor[gray]{1}\textbf{what} &     \cellcolor[gray]{1}\textbf{med} & \cellcolor[gray]{1}\textbf{IQR} & \cellcolor[gray]{1}\\
            \hline
            \rowcolor{white}1 &      A1 &    0.50 &  0.26 & \quart{37}{26}{50}{100} \\
            \rowcolor{blue!10}\textbf{2} &      \textbf{D1} &    \textbf{0.74} &  \textbf{0.35} & \quart{53}{35}{74}{100} \\
            \rowcolor{white}3 &      C2 &    1.00 &  0.01 & \quart{99}{1}{100}{100} \\
            \rowcolor{white}3 &      B4 &    1.00 &  0.01 & \quart{99}{1}{100}{100} \\
            \rowcolor{white}3 &      B2 &    1.00 &  0.01 & \quart{99}{1}{100}{100} \\
            \rowcolor{red!10}\textbf{3} &      \textbf{B1} &    \textbf{1.00} &  \textbf{0.00} & \quart{100}{0}{100}{100} \\
            \rowcolor{red!10}\textbf{3} &      \textbf{B3} &    \textbf{1.00} &  \textbf{0.00} & \quart{100}{0}{100}{100} \\
            \rowcolor{white}3 &      A2 &    1.00 &  0.00 & \quart{100}{0}{100}{100} \\
        \end{tabular} &
        \\
         & (j). Project Name: galaxyProject/galaxy & \\
    \end{tabular}
  \end{adjustbox}
   \end{center}
\end{table*}

\begin{table*}[!t]
    \caption{Scott-Knott analysis results from 20 open-source software engineering projects.
    In these tables, \colorbox{blue!10}{blue row} marks the performance of D1 algorithm, while \colorbox{red!10}{red row} denotes the performance of B1/B3 approaches.
    Note that in 13/20 of these results,
    B1/B3 is ranked the same as the omniscient A2 method: see figures b,c,d,e,f,g,h,i,m,o,q,s,t. \revised{The algorithm} with n/a \revised{means} they are too expensive to finish so that they are in the lowest rank.
    }
    \label{tbl:opensourceResultSE}
    \centering
    \begin{adjustbox}{max width=.85\textwidth}
    \scriptsize
    
    \begin{tabular}{c@{}c@{}c}
        \begin{tabular}{ccccl}
            \cellcolor[gray]{1}\textbf{rank} & \cellcolor[gray]{1}\textbf{what} &     \cellcolor[gray]{1}\textbf{med} & \cellcolor[gray]{1}\textbf{IQR} & \cellcolor[gray]{1}\\
            \hline
            \rowcolor{white}1 &      A1 &    0.52 &  0.28 & \quart{41}{28}{52}{100} \\
            \rowcolor{blue!10}\textbf{1} &      \textbf{D1} &    \textbf{0.59} &  \textbf{0.42} & \quart{33}{42}{59}{100} \\
            \rowcolor{white}2 &      C2 &    0.89 &  0.32 & \quart{66}{32}{89}{100} \\
            \rowcolor{white}2 &      C1 &    0.91 &  0.18 & \quart{79}{18}{91}{100} \\
            \rowcolor{white}2 &      B2 &    0.91 &  0.23 & \quart{74}{23}{91}{100} \\
            \rowcolor{white}2 &      B4 &    0.95 &  0.22 & \quart{76}{22}{95}{100} \\
            \rowcolor{red!10}\textbf{3} &      \textbf{B1} &    \textbf{0.97} &  \textbf{0.17} & \quart{82}{17}{97}{100} \\
            \rowcolor{red!10}\textbf{3} &      \textbf{B3} &    \textbf{0.97} &  \textbf{0.14} & \quart{85}{14}{97}{100} \\
            \rowcolor{white}4 &      A2 &    0.99 &  0.02 & \quart{97}{2}{99}{100} \\
        \end{tabular} & 
        \begin{tabular}{|ccccl}
            \cellcolor[gray]{1}\textbf{rank} & \cellcolor[gray]{1}\textbf{what} &     \cellcolor[gray]{1}\textbf{med} & \cellcolor[gray]{1}\textbf{IQR} & \cellcolor[gray]{1}\\
            \hline
            \rowcolor{white}1 &      A1 &    0.48 &  0.31 & \quart{31}{31}{48}{100} \\
            \rowcolor{blue!10}\textbf{2} &      \textbf{D1} &    \textbf{0.65} &  \textbf{0.32} & \quart{46}{32}{65}{100} \\
            \rowcolor{white}3 &      C2 &    0.75 &  0.25 & \quart{62}{25}{75}{100} \\
            \rowcolor{white}4 &      C1 &    0.94 &  0.14 & \quart{82}{14}{94}{100} \\
            \rowcolor{white}4 &      B2 &    0.95 &  0.20 & \quart{78}{20}{95}{100} \\
            \rowcolor{white}5 &      B4 &    0.97 &  0.16 & \quart{82}{16}{97}{100} \\
            \rowcolor{red!10}\textbf{5} &      \textbf{B1} &    \textbf{0.98} &  \textbf{0.09} & \quart{89}{9}{98}{100} \\
            \rowcolor{red!10}\textbf{5} &      \textbf{B3} &    \textbf{0.98} &  \textbf{0.06} & \quart{92}{6}{98}{100} \\
            \rowcolor{white}5 &      A2 &    0.98 &  0.00 & \quart{98}{0}{98}{100} \\
        \end{tabular} &
        \begin{tabular}{|ccccl}
            \cellcolor[gray]{1}\textbf{rank} & \cellcolor[gray]{1}\textbf{what} &     \cellcolor[gray]{1}\textbf{med} & \cellcolor[gray]{1}\textbf{IQR} & \cellcolor[gray]{1}\\
            \hline
            \rowcolor{white}1 &      A1 &    0.49 &  0.23 & \quart{37}{23}{49}{100} \\
            \rowcolor{blue!10}\textbf{2} &      \textbf{D1} &    \textbf{0.67} &  \textbf{0.37} & \quart{45}{37}{67}{100} \\
            \rowcolor{white}3 &      C2 &    0.78 &  0.28 & \quart{62}{28}{78}{100} \\
            \rowcolor{white}4 &      B2 &    0.92 &  0.19 & \quart{78}{19}{92}{100} \\
            \rowcolor{white}4 &      C1 &    0.93 &  0.14 & \quart{82}{14}{93}{100} \\
            \rowcolor{white}4 &      B4 &    0.94 &  0.14 & \quart{84}{14}{94}{100} \\
            \rowcolor{red!10}\textbf{5}  &      \textbf{B1} &    \textbf{0.98} &  \textbf{0.04} & \quart{95}{4}{98}{100} \\
            \rowcolor{red!10}\textbf{5} &      \textbf{B3} &    \textbf{0.98} &  \textbf{0.03} & \quart{96}{3}{98}{100} \\
            \rowcolor{white}5 &      A2 &    0.99 &  0.01 & \quart{98}{1}{99}{100} \\
        \end{tabular} \\
        (a). Project Name: loomio/loomio & (b). Project Name: languagetool-org/languagetool & (c). Project Name: deeplearning4j/deeplearning4j 
        \\
        \\
        \begin{tabular}{ccccl}
            \cellcolor[gray]{1}\textbf{rank} & \cellcolor[gray]{1}\textbf{what} &     \cellcolor[gray]{1}\textbf{med} & \cellcolor[gray]{1}\textbf{IQR} & \cellcolor[gray]{1}\\
            \hline
            \rowcolor{white}1 &      A1 &    0.51 &  0.23 & \quart{39}{23}{51}{100} \\
            \rowcolor{blue!10}\textbf{2} &      \textbf{D1} &    \textbf{0.67} &  \textbf{0.26} & \quart{52}{26}{67}{100} \\
            \rowcolor{white}3 &      C2 &    0.78 &  0.22 & \quart{67}{22}{78}{100} \\
            \rowcolor{white}4 &      C1 &    0.92 &  0.10 & \quart{84}{10}{92}{100} \\
            \rowcolor{white}4 &      B2 &    0.93 &  0.10 & \quart{86}{10}{93}{100} \\
            \rowcolor{white}4 &      B4 &    0.94 &  0.09 & \quart{87}{9}{94}{100} \\
            \rowcolor{red!10}\textbf{5} &      \textbf{B1} &    \textbf{0.96} &  \textbf{0.04} & \quart{93}{4}{96}{100} \\
            \rowcolor{red!10}\textbf{5} &      \textbf{B3} &    \textbf{0.96} &  \textbf{0.03} & \quart{94}{3}{96}{100} \\
            \rowcolor{white}5 &      A2 &    0.97 &  0.01 & \quart{96}{1}{97}{100} \\
        \end{tabular} &
        \begin{tabular}{|ccccl}
            \cellcolor[gray]{1}\textbf{rank} & \cellcolor[gray]{1}\textbf{what} &     \cellcolor[gray]{1}\textbf{med} & \cellcolor[gray]{1}\textbf{IQR} & \cellcolor[gray]{1}\\
            \hline
            \rowcolor{white}1 &      A1 &    0.49 &  0.24 & \quart{36}{24}{49}{100} \\
            \rowcolor{blue!10}\textbf{2} &      \textbf{D1} &    \textbf{0.68} &  \textbf{0.37} & \quart{48}{37}{68}{100} \\
            \rowcolor{white}2 &      C2 &    0.68 &  0.36 & \quart{57}{36}{68}{100} \\
            \rowcolor{white}3 &      B2 &    0.96 &  0.16 & \quart{83}{16}{96}{100} \\
            \rowcolor{white}3 &      C1 &    0.97 &  0.09 & \quart{90}{9}{97}{100} \\
            \rowcolor{white}3 &      B4 &    0.98 &  0.12 & \quart{87}{12}{98}{100} \\
            \rowcolor{red!10}\textbf{4} &      \textbf{B1} &    \textbf{0.99} &  \textbf{0.01} & \quart{99}{1}{99}{100} \\
            \rowcolor{red!10}\textbf{4} &      \textbf{B3} &    \textbf{0.99} &  \textbf{0.02} & \quart{98}{2}{99}{100} \\
            \rowcolor{white}4 &      A2 &    1.00 &  0.01 & \quart{99}{1}{100}{100} \\
        \end{tabular} & 
        \begin{tabular}{|ccccl}
            \cellcolor[gray]{1}\textbf{rank} & \cellcolor[gray]{1}\textbf{what} &     \cellcolor[gray]{1}\textbf{med} & \cellcolor[gray]{1}\textbf{IQR} & \cellcolor[gray]{1}\\
            \hline
            \rowcolor{white}1 &      A1 &    0.50 &  0.22 & \quart{39}{22}{50}{100} \\
            \rowcolor{blue!10}\textbf{2} &      \textbf{D1} &    \textbf{0.73} &  \textbf{0.35} & \quart{51}{35}{73}{100} \\
            \rowcolor{white}3 &      C2 &    0.78 &  0.24 & \quart{65}{24}{78}{100} \\
            \rowcolor{white}4 &      C1 &    0.97 &  0.04 & \quart{94}{4}{97}{100} \\
            \rowcolor{white}4 &      B2 &    0.97 &  0.09 & \quart{90}{9}{97}{100} \\
            \rowcolor{white}4 &      B4 &    0.97 &  0.07 & \quart{92}{7}{97}{100} \\
            \rowcolor{red!10}\textbf{5} &      \textbf{B1} &    \textbf{0.99} &  \textbf{0.02} & \quart{97}{2}{99}{100} \\
            \rowcolor{red!10}\textbf{5} &      \textbf{B3} &    \textbf{0.99} &  \textbf{0.02} & \quart{97}{2}{99}{100} \\
            \rowcolor{white}5 &      A2 &    0.99 &  0.00 & \quart{99}{0}{99}{100} \\
        \end{tabular} \\
        (d). Project Name: Unidata/thredds & (e). Project Name: nutzam/nutz & (f). Project Name: structr/structr
        \\
        \\
        \begin{tabular}{ccccl}
            \cellcolor[gray]{1}\textbf{rank} & \cellcolor[gray]{1}\textbf{what} &     \cellcolor[gray]{1}\textbf{med} & \cellcolor[gray]{1}\textbf{IQR} & \cellcolor[gray]{1}\\
            \hline
            \rowcolor{white}1 &      A1 &    0.47 &  0.24 & \quart{37}{24}{47}{100} \\
            \rowcolor{blue!10}\textbf{2} &      \textbf{D1} &    \textbf{0.63} &  \textbf{0.34} & \quart{41}{34}{63}{100} \\
            \rowcolor{white}3 &      C2 &    0.72 &  0.24 & \quart{61}{24}{72}{100} \\
            \rowcolor{white}4 &      B2 &    0.94 &  0.24 & \quart{74}{24}{94}{100} \\
            \rowcolor{white}4 &      C1 &    0.95 &  0.17 & \quart{81}{17}{95}{100} \\
            \rowcolor{white}4 &      B4 &    0.96 &  0.24 & \quart{74}{24}{96}{100} \\
            \rowcolor{red!10}\textbf{5} &      \textbf{B1} &    \textbf{0.98} &  \textbf{0.13} & \quart{86}{13}{98}{100} \\
            \rowcolor{red!10}\textbf{5} &      \textbf{B3} &    \textbf{0.98} &  \textbf{0.10} & \quart{89}{10}{98}{100} \\
            \rowcolor{white}5 &      A2 &    0.99 &  0.01 & \quart{98}{1}{99}{100} \\
        \end{tabular} & 
        \begin{tabular}{|ccccl}
            \cellcolor[gray]{1}\textbf{rank} & \cellcolor[gray]{1}\textbf{what} &     \cellcolor[gray]{1}\textbf{med} & \cellcolor[gray]{1}\textbf{IQR} & \cellcolor[gray]{1}\\
            \hline
            \rowcolor{white}1 &      A1 &    0.51 &  0.23 & \quart{39}{23}{51}{100} \\
            \rowcolor{blue!10}\textbf{2} &      \textbf{D1} &    \textbf{0.74} &  \textbf{0.29} & \quart{55}{29}{74}{100} \\
            \rowcolor{white}3 &      C2 &    0.80 &  0.20 & \quart{70}{20}{80}{100} \\
            \rowcolor{white}4 &      C1 &    0.98 &  0.02 & \quart{96}{2}{98}{100} \\
            \rowcolor{white}4 &      B4 &    0.98 &  0.03 & \quart{96}{3}{98}{100} \\
            \rowcolor{white}4 &      B2 &    0.98 &  0.04 & \quart{95}{4}{98}{100} \\
            \rowcolor{red!10}\textbf{4} &      \textbf{B1} &    \textbf{0.98} &  \textbf{0.07} & \quart{92}{7}{98}{100} \\
            \rowcolor{red!10}\textbf{4} &      \textbf{B3} &    \textbf{0.98} &  \textbf{0.06} & \quart{93}{6}{98}{100} \\
            \rowcolor{white}4 &      A2 &    0.99 &  0.00 & \quart{99}{0}{99}{100} \\
        \end{tabular} &
        \begin{tabular}{|ccccl}
            \cellcolor[gray]{1}\textbf{rank} & \cellcolor[gray]{1}\textbf{what} &     \cellcolor[gray]{1}\textbf{med} & \cellcolor[gray]{1}\textbf{IQR} & \cellcolor[gray]{1}\\
            \hline
            \rowcolor{white}1 &      A1 &    0.51 &  0.28 & \quart{37}{28}{51}{100} \\
            \rowcolor{blue!10}\textbf{2} &      \textbf{D1} &    \textbf{0.63} &  \textbf{0.36} & \quart{44}{36}{63}{100} \\
            \rowcolor{white}3 &      C2 &    0.72 &  0.23 & \quart{62}{23}{72}{100} \\
            \rowcolor{white}4 &      C1 &    0.96 &  0.03 & \quart{94}{3}{96}{100} \\
            \rowcolor{white}4 &      B2 &    0.97 &  0.04 & \quart{94}{4}{97}{100} \\
            \rowcolor{white}4 &      B4 &    0.98 &  0.03 & \quart{95}{3}{98}{100} \\
            \rowcolor{red!10}\textbf{4} &      \textbf{B1} &    \textbf{0.98} &  \textbf{0.01} & \quart{97}{1}{98}{100} \\
            \rowcolor{red!10}\textbf{4} &      \textbf{B3} &    \textbf{0.98} &  \textbf{0.00} & \quart{98}{0}{98}{100} \\
            \rowcolor{white}4 &      A2 &    0.98 &  0.00 & \quart{98}{0}{98}{100} \\
        \end{tabular} \\
        (g). Project Name: ocpsoft/rewrite & (h). Project Name: eclipse/jetty.project & (i): Project Name: square/okhttp
        \\
        \\
        \begin{tabular}{ccccl}
            \cellcolor[gray]{1}\textbf{rank} & \cellcolor[gray]{1}\textbf{what} &     \cellcolor[gray]{1}\textbf{med} & \cellcolor[gray]{1}\textbf{IQR} & \cellcolor[gray]{1}\\
            \hline
            \rowcolor{white}1 &      A1 &    0.53 &  0.34 & \quart{35}{34}{53}{100} \\
            \rowcolor{blue!10}\textbf{1} &      \textbf{D1} &    \textbf{0.59} &  \textbf{0.46} & \quart{32}{46}{59}{100} \\
            \rowcolor{white}2 &      B2 &    0.88 &  0.22 & \quart{71}{22}{88}{100} \\
            \rowcolor{white}2 &      C2 &    0.88 &  0.36 & \quart{56}{36}{88}{100} \\
            \rowcolor{white}2 &      B4 &    0.89 &  0.28 & \quart{68}{28}{89}{100} \\
            \rowcolor{white}2 &      C1 &    0.90 &  0.15 & \quart{81}{15}{90}{100} \\
            \rowcolor{red!10}\textbf{3} &      \textbf{B1} &    \textbf{0.96} &  \textbf{0.15} & \quart{84}{15}{96}{100} \\
            \rowcolor{red!10}\textbf{3} &      \textbf{B3} &    \textbf{0.96} &  \textbf{0.15} & \quart{84}{15}{96}{100} \\
            \rowcolor{white}4 &      A2 &    0.99 &  0.00 & \quart{99}{0}{99}{100} \\
        \end{tabular} &
        \begin{tabular}{|ccccl}
            \cellcolor[gray]{1}\textbf{rank} & \cellcolor[gray]{1}\textbf{what} &     \cellcolor[gray]{1}\textbf{med} & \cellcolor[gray]{1}\textbf{IQR} & \cellcolor[gray]{1}\\
            \hline
            \rowcolor{white}1 &      A1 &    0.49 &  0.19 & \quart{36}{19}{49}{100} \\
            \rowcolor{blue!10}\textbf{2} &      \textbf{D1} &    \textbf{0.63} &  \textbf{0.32} & \quart{46}{32}{63}{100} \\
            \rowcolor{white}3 &      B2 &    0.71 &  0.31 & \quart{59}{31}{71}{100} \\
            \rowcolor{white}3 &      B4 &    0.72 &  0.30 & \quart{61}{30}{72}{100} \\
            \rowcolor{white}3 &      C1 &    0.72 &  0.34 & \quart{61}{34}{72}{100} \\
            \rowcolor{white}3 &      C2 &    0.73 &  0.19 & \quart{66}{19}{73}{100} \\
            \rowcolor{red!10}\textbf{4} &      \textbf{B1} &    \textbf{0.93} &  \textbf{0.26} & \quart{71}{26}{93}{100} \\
            \rowcolor{red!10}\textbf{4} &      \textbf{B3} &    \textbf{0.94} &  \textbf{0.27} & \quart{71}{27}{94}{100} \\
            \rowcolor{white}5 &      A2 &    0.99 &  0.2 & \quart{97}{2}{99}{100} \\
        \end{tabular} &
        \begin{tabular}{|ccccl}
            \cellcolor[gray]{1}\textbf{rank} & \cellcolor[gray]{1}\textbf{what} &     \cellcolor[gray]{1}\textbf{med} & \cellcolor[gray]{1}\textbf{IQR} & \cellcolor[gray]{1}\\
            \hline
            \rowcolor{white}1 &      A1 &    0.50 &  0.24 & \quart{38}{24}{50}{100} \\
            \rowcolor{blue!10}\textbf{2} &      \textbf{D1} &    \textbf{0.69} &  \textbf{0.37} & \quart{42}{37}{69}{100} \\
            \rowcolor{white}3 &      C2 &    0.70 &  0.26 & \quart{57}{26}{70}{100} \\
            \rowcolor{white}4 &      C1 &    0.91 &  0.21 & \quart{74}{21}{91}{100} \\
            \rowcolor{white}4 &      B2 &    0.91 &  0.24 & \quart{72}{24}{91}{100} \\
            \rowcolor{white}4 &      B4 &    0.92 &  0.23 & \quart{73}{23}{92}{100} \\
            \rowcolor{red!10}\textbf{5} &      \textbf{B1} &    \textbf{0.96} &  \textbf{0.12} & \quart{87}{12}{96}{100} \\
            \rowcolor{red!10}\textbf{5} &      \textbf{B3} &    \textbf{0.96} &  \textbf{0.12} & \quart{87}{12}{96}{100} \\
            \rowcolor{white}6 &      A2 &    0.99 &  0.03 & \quart{96}{3}{99}{100} \\
        \end{tabular} \\
         (j). Project Name: openSUSE/open-build-service & (k). Project Name: thinkaurelius/titan & (l): Project Name: Graylog2/graylog2-server
        \\
        \\
        \begin{tabular}{ccccl}
            \cellcolor[gray]{1}\textbf{rank} & \cellcolor[gray]{1}\textbf{what} &     \cellcolor[gray]{1}\textbf{med} & \cellcolor[gray]{1}\textbf{IQR} & \cellcolor[gray]{1}\\
            \hline
            \rowcolor{white}1 &      A1 &    0.48 &  0.32 & \quart{32}{32}{48}{100} \\
            \rowcolor{blue!10}\textbf{2} &      \textbf{D1} &    \textbf{0.66} &  \textbf{0.36} & \quart{45}{36}{66}{100} \\
            \rowcolor{white}3 &      C2 &    0.95 &  0.28 & \quart{71}{28}{95}{100} \\
            \rowcolor{white}4 &      B4 &    0.98 &  0.03 & \quart{96}{3}{98}{100} \\
            \rowcolor{red!10}\textbf{4} &      \textbf{B1} &    \textbf{0.98} &  \textbf{0.03} & \quart{96}{3}{98}{100} \\
            \rowcolor{white}4 &      B2 &    0.98 &  0.03 & \quart{96}{3}{98}{100} \\
            \rowcolor{red!10}\textbf{4} &      \textbf{B3} &    \textbf{0.98} &  \textbf{0.03} & \quart{96}{3}{98}{100} \\
            \rowcolor{white}4 &      C1 &    0.98 &  0.03 & \quart{96}{3}{98}{100} \\
            \rowcolor{white}4 &      A2 &    0.98 &  0.01 & \quart{98}{1}{98}{100} \\
        \end{tabular} & 
        \begin{tabular}{|ccccl}
            \cellcolor[gray]{1}\textbf{rank} & \cellcolor[gray]{1}\textbf{what} &     \cellcolor[gray]{1}\textbf{med} & \cellcolor[gray]{1}\textbf{IQR} & \cellcolor[gray]{1}\\
            \hline
            \rowcolor{white}1 &      A1 &    0.51 &  0.31 & \quart{34}{31}{51}{100} \\
            \rowcolor{blue!10}\textbf{2} &      \textbf{D1} &    \textbf{0.64} &  \textbf{0.41} & \quart{40}{41}{64}{100} \\
            \rowcolor{white}3 &      B2 &    0.83 &  0.28 & \quart{65}{28}{83}{100} \\
            \rowcolor{white}3 &      C2 &    0.84 &  0.24 & \quart{68}{24}{84}{100} \\
            \rowcolor{white}3 &      B4 &    0.85 &  0.24 & \quart{70}{24}{85}{100} \\
            \rowcolor{white}3 &      C1 &    0.88 &  0.25 & \quart{72}{25}{88}{100} \\
            \rowcolor{red!10}\textbf{4} &      \textbf{B1} &    \textbf{0.98} &  \textbf{0.09} & \quart{91}{9}{98}{100} \\
            \rowcolor{red!10}\textbf{4} &      \textbf{B3} &    \textbf{0.98} &  \textbf{0.09} & \quart{91}{9}{98}{100} \\
            \rowcolor{white}5 &      A2 &    1.00 &  0.02 & \quart{98}{2}{100}{100} \\
        \end{tabular} &
        \begin{tabular}{|ccccl}
            \cellcolor[gray]{1}\textbf{rank} & \cellcolor[gray]{1}\textbf{what} &     \cellcolor[gray]{1}\textbf{med} & \cellcolor[gray]{1}\textbf{IQR} & \cellcolor[gray]{1}\\
            \hline
            \rowcolor{white}1 &      A1 &    0.48 &  0.45 & \quart{26}{45}{48}{100} \\
            \rowcolor{blue!10}\textbf{1} &      \textbf{D1} &    \textbf{0.51} &  \textbf{0.38} & \quart{31}{38}{51}{100} \\
            \rowcolor{white}2 &      B2 &    0.96 &  0.33 & \quart{66}{33}{96}{100} \\
            \rowcolor{white}2 &      C1 &    0.96 &  0.15 & \quart{82}{15}{96}{100} \\
            \rowcolor{white}2 &      B4 &    0.97 &  0.29 & \quart{70}{29}{97}{100} \\
            \rowcolor{white}2 &      C2 &    0.97 &  0.30 & \quart{69}{30}{97}{100} \\
            \rowcolor{red!10}\textbf{3} &      \textbf{B1} &    \textbf{0.99} &  \textbf{0.03} & \quart{96}{3}{99}{100} \\
            \rowcolor{red!10}\textbf{3} &      \textbf{B3} &    \textbf{0.99} &  \textbf{0.03} & \quart{96}{3}{99}{100} \\
            \rowcolor{white}3 &      A2 &    0.99 &  0.00 & \quart{99}{0}{99}{100} \\
        \end{tabular} \\
        (m). Project Name: puppetlabs/puppet & (n). Project Name: middleman/middleman & (o): Project Name: locomotivecms/engine
        \\
        \\
        \begin{tabular}{ccccl}
            \cellcolor[gray]{1}\textbf{rank} & \cellcolor[gray]{1}\textbf{what} &     \cellcolor[gray]{1}\textbf{med} & \cellcolor[gray]{1}\textbf{IQR} & \cellcolor[gray]{1}\\
            \hline
            \rowcolor{white}1 &      A1 &    0.50 &  0.24 & \quart{38}{24}{50}{100} \\
            \rowcolor{blue!10}\textbf{2} &      \textbf{D1} &    \textbf{0.72} &  \textbf{0.39} & \quart{42}{39}{72}{100} \\
            \rowcolor{white}3 &      B2 &    0.78 &  0.36 & \quart{55}{36}{78}{100} \\
            \rowcolor{white}3 &      B4 &    0.79 &  0.37 & \quart{56}{37}{79}{100} \\
            \rowcolor{white}3 &      C2 &    0.81 &  0.33 & \quart{59}{33}{81}{100} \\
            \rowcolor{white}3 &      C1 &    0.83 &  0.32 & \quart{61}{32}{83}{100} \\
            \rowcolor{red!10}\textbf{4} &      \textbf{B1} &    \textbf{0.90} &  \textbf{0.17} & \quart{82}{17}{90}{100} \\
            \rowcolor{red!10}\textbf{4} &      \textbf{B3} &    \textbf{0.94} &  \textbf{0.15} & \quart{84}{15}{94}{100} \\
            \rowcolor{white}5 &      A2 &    0.99 &  0.08 & \quart{92}{8}{99}{100} \\
        \end{tabular} &
        \begin{tabular}{|ccccl}
            \cellcolor[gray]{1}\textbf{rank} & \cellcolor[gray]{1}\textbf{what} &     \cellcolor[gray]{1}\textbf{med} & \cellcolor[gray]{1}\textbf{IQR} & \cellcolor[gray]{1}\\
            \hline
            \rowcolor{white}1 &      A1 &    0.50 &  0.25 & \quart{37}{25}{50}{100} \\
            \rowcolor{blue!10}\textbf{2} &      \textbf{D1} &    \textbf{0.67} &  \textbf{0.41} & \quart{44}{41}{67}{100} \\
            \rowcolor{white}3 &      C2 &    0.75 &  0.24 & \quart{63}{24}{75}{100} \\
            \rowcolor{white}4 &      B2 &    0.97 &  0.10 & \quart{90}{10}{97}{100} \\
            \rowcolor{white}4 &      B4 &    0.97 &  0.10 & \quart{90}{10}{97}{100} \\
            \rowcolor{white}5 &      C1 &    1.00 &  0.03 & \quart{97}{3}{100}{100} \\
            \rowcolor{red!10}\textbf{5} &      \textbf{B1} &    \textbf{1.00} &  \textbf{0.01} & \quart{100}{1}{100}{100} \\
            \rowcolor{red!10}\textbf{5} &      \textbf{B3} &    \textbf{1.00} &  \textbf{0.01} & \quart{100}{1}{100}{100} \\
            \rowcolor{white}5 &      A2 &    1.00 &  0.00 & \quart{100}{0}{100}{100} \\
        \end{tabular} &
        \begin{tabular}{|ccccl}
            \cellcolor[gray]{1}\textbf{rank} & \cellcolor[gray]{1}\textbf{what} &     \cellcolor[gray]{1}\textbf{med} & \cellcolor[gray]{1}\textbf{IQR} & \cellcolor[gray]{1}\\
            \hline
            \rowcolor{white}1 &      A1 &    0.52 &  0.23 & \quart{45}{23}{52}{100} \\
            \rowcolor{blue!10}\textbf{2} &      \textbf{D1} &    \textbf{0.70} &  \textbf{0.28} & \quart{48}{28}{70}{100} \\
            \rowcolor{white}3 &      C2 &    0.73 &  0.32 & \quart{66}{32}{73}{100} \\
            \rowcolor{white}4 &      C1 &    0.82 &  0.16 & \quart{79}{16}{82}{100} \\
            \rowcolor{white}4 &      B2 &    0.83 &  0.17 & \quart{80}{17}{83}{100} \\
            \rowcolor{white}4 &      B4 &    0.83 &  0.17 & \quart{80}{17}{83}{100} \\
            \rowcolor{red!10}\textbf{5} &      \textbf{B1} &    \textbf{0.86} &  \textbf{0.19} & \quart{80}{19}{86}{100} \\
            \rowcolor{red!10}\textbf{5} &      \textbf{B3} &    \textbf{0.86} &  \textbf{0.17} & \quart{82}{17}{86}{100} \\
            \rowcolor{white}6 &      A2 &    0.98 &  0.15 & \quart{84}{15}{98}{100} \\
        \end{tabular} \\
        (p). Project Name: diaspora/diaspora & (q). Project Name: facebook/presto & (r): Project Name: rspec/rspec-core
        \\
        \\
        \begin{tabular}{ccccl}
            \cellcolor[gray]{1}\textbf{rank} & \cellcolor[gray]{1}\textbf{what} &     \cellcolor[gray]{1}\textbf{med} & \cellcolor[gray]{1}\textbf{IQR} & \cellcolor[gray]{1}\\
            \hline
            \rowcolor{white}1 &      A1 &    0.50 &  0.12 & \quart{44}{12}{50}{100} \\
            \rowcolor{white}1 &      C1 &    n/a  &  n/a  & \\
            \rowcolor{white}2 &      C2 &    0.86 &  0.11 & \quart{78}{11}{86}{100} \\
            \rowcolor{blue!10}\textbf{2} &      \textbf{D1} &    \textbf{0.88} &  \textbf{0.17} & \quart{75}{17}{88}{100} \\
            \rowcolor{white}2 &      B2 &    0.88 &  0.14 & \quart{78}{14}{88}{100} \\
            \rowcolor{white}2 &      B4 &    0.88 &  0.15 & \quart{78}{15}{88}{100} \\
            \rowcolor{red!10}\textbf{3} &      \textbf{B1} &    \textbf{0.98} &  \textbf{0.02} & \quart{97}{2}{98}{100} \\
            \rowcolor{red!10}\textbf{3} &      \textbf{B3} &    \textbf{0.98} &  \textbf{0.02} & \quart{97}{2}{98}{100} \\
            \rowcolor{white}3 &      A2 &    0.99 &  0.02 & \quart{98}{2}{99}{100} \\
        \end{tabular} & 
        \begin{tabular}{|ccccl}
            \cellcolor[gray]{1}\textbf{rank} & \cellcolor[gray]{1}\textbf{what} &     \cellcolor[gray]{1}\textbf{med} & \cellcolor[gray]{1}\textbf{IQR} & \cellcolor[gray]{1}\\
            \hline
            \rowcolor{white}1 &      A1 &    0.50 &  0.04 & \quart{48}{4}{50}{100} \\
            \rowcolor{blue!10}\textbf{1} &      \textbf{D1} &    \textbf{n/a} & \textbf{n/a} & \\
            \rowcolor{white}1 &      C1 &    n/a & n/a & \\
            \rowcolor{white}1 &      C2 &    n/a & n/a & \\
            \rowcolor{white}2 &      B2 &    0.97 &  0.07 & \quart{92}{7}{97}{100} \\
            \rowcolor{white}2 &      B4 &    0.97 &  0.07 & \quart{92}{7}{97}{100} \\
            \rowcolor{red!10}\textbf{3} &      \textbf{B1} &    \textbf{0.99} &  \textbf{0.04} & \quart{96}{4}{99}{100} \\
            \rowcolor{red!10}\textbf{3} &      \textbf{B3} &    \textbf{0.99} &  \textbf{0.04} & \quart{96}{4}{99}{100} \\
            \rowcolor{white}3 &      A2 &    0.99 &  0.01 & \quart{99}{1}{99}{100} \\
        \end{tabular} &
        \\
        (s). Project Name: rails/rails & (t). Project Name: jruby/jruby
    \end{tabular}
    \end{adjustbox}
\end{table*}

As seen in~\tbl{SK_LN}, as might be expected, the performance \revised{of} all \revised{the} algorithms are bounded by the dumbest  A1 prioritization algorithm (which performed worse) and the \revised{most} omniscient A2 algorithm (that performed best).  


\begin{figure}[!b]
    \centering
    \includegraphics[width=0.47\textwidth]{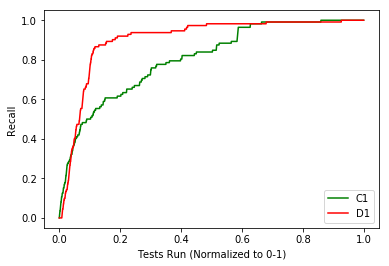}
    \caption{Mean fault detection rates. X-axis = number of tests executed, Y-axis = "recall" (percentage of   failing
    test suites).   }
    \label{fig:Rate_LN}
\end{figure}

After A2, we see that D1 and C1 are tied \revised{in} the best place (in rank 5). That \revised{says}, we recommend D1 over C1 \revised{due to the following reasons}:
\bi
    \item D1 runs five times faster than C1 (in our proprietary project, 328 seconds versus 1457 seconds).
    \item D1 converges faster to a higher plateau of performance. \fig{Rate_LN} records the percentage of failed test cases that are explored with increasing number of tests that are executed in the last test run. In~\fig{Rate_LN}, X-axis represents the percentage of tests that are executed, and Y-axis represents the percentage of failing test cases that are explored. We say {\em D1 converges faster} since D1 explores 85\% failed test cases while C1 only explores 55\% when 15\% tests are executed.
\ei

Hence we answer {\bf RQ1} as follows:
\begin{blockquote}
    \noindent
    {\em The {\em D1} prioritization scheme, which is {\em TERMINATOR}}, works best for that closed-source project.
\end{blockquote}

\subsection{What is the best algorithm for open-source projects? (RQ2)} \label{RQ2}

\revised{To} answer RQ2, we use 10 computational science (CS) projects and 20 software engineering (SE) projects from GitHub.~\tbl{opensourceResultCS} shows the Scott-Knott analysis for 10 CS projects and~\tbl{opensourceResultSE} states the results for 20 SE projects. From comparisons among all 30 projects, we observe that for all these open-source projects, B1 and B3 always perform better than any other algorithms.
\revised{Interestingly, B1 and B3 schemes rank in the same level} as the omniscient A2 algorithm \revised{does} in 8/10 of the~\tbl{opensourceResultCS} results and 13/21 of the~\tbl{opensourceResultSE} results. 
That is, in the majority case, B1 and B3 are performing in such a high level where they cannot be beaten.

Moreover, in our experiments, we find C1 takes a very long time in prioritizing projects which have over 800 test builds or 1500 failed test cases. For example, in the \textit{Reaction Mechanism Generator} project, which has 850 test builds and 617 failed test cases, C1 takes around 48 hours to simulate 70\% test builds. Therefore, we conclude that C1 is a very computational costly algorithm which has issues scaling up to projects with a huge number of test builds or failed test cases. C1 performs so slowly that we do not use it for our analysis of  projects with more than 800 test builds or more than 1500 failed test cases.

In summary, we can conduct the answer for RQ2 based on the above results:
\begin{blockquote}
    \noindent
    \em{For open-source projects, the best approach is not {\em D1 TERMINATOR}, but rather to prioritize using either {\em B1}, which is {\em passing times  since  last  failure}  or  {\em B3}, which implements {\em exponential  metric}.
    }
\end{blockquote}

\subsection{Are different prioritization algorithms perform various in the open-source projects and the closed-source project? (RQ3)} \label{RQ3}
To answer RQ3, we look at the B1/B3 and D1 results in~\tbl{SK_LN},~\tbl{opensourceResultCS}, and~\tbl{opensourceResultSE}. We highlight D1 result with 
\colorbox{blue!10}{blue}
 and 
 the B1/B3 results with \colorbox{red!10}{red}. Note that the ranking of these algorithms is reversed for our closed-source and open-source examples:
\bi
    \item As shown in~\tbl{SK_LN}, for our close-sourced case study, D1 \revised{is} seen to perform much better than B1/B3.
    \item However, as shown in~\tbl{opensourceResultCS} and~\tbl{opensourceResultSE}, for open-source projects, that ranking is completely reverse,
\ei
Based on these points, we can answer RQ3 that
\begin{blockquote}
    \noindent
    \em{Test  case  prioritization  schemes  that  work best for the industrial closed-source project can work worse for open-source projects (and vice versa)}
\end{blockquote}

\section{Discussion} \label{Discussion} 

\begin{table*}[b!]
    \centering
    \caption{Run time for all algorithms (unit: (s)). \colorbox{gray!115}{\color{white}Dark gray} marks the performance of D1 in the proprietary closed-source project from our industrial partner, which is an acceptable run time. \colorbox{gray!25}{Light gray} marks the performances of B1 and B3 in the open-source projects, which are much shorter than D1.}
    \label{tbl:runtime}
    \begin{adjustbox}{max width=\textwidth}
    \scriptsize
    \begin{tabular}{c@{}c}
        \begin{tabular}{l|r|r|r|r|r|r|r}
            \rowcolor{gray!1}
            \textbf{Project Name} & \textbf{B1} & \textbf{B2} & \textbf{B3} & \textbf{B4} & \textbf{C1} & \textbf{C2} & \textbf{D1} \\
            \hline
            \rowcolor[HTML]{FFFFFF} Unidata/thredds & \cellcolor{gray!25}0.2 & 0.3 & \cellcolor{gray!25}0.3 & 0.4 & 8.4 & 1.2 & 1.7 \\
            
            \rowcolor[HTML]{F3F3F3} OpenSUSE & \cellcolor{gray!25}0.2 & 0.3 & \cellcolor{gray!25} 0.3 & 0.3 & 11.4 & 1.1 & 1.9 \\
            
            \rowcolor[HTML]{FFFFFF} Thinkaurelius & \cellcolor{gray!25}0.2 & 0.3 & \cellcolor{gray!25} 0.3 & 0.4 & 16.8 & 2.7 & 2.9 \\
            
            \rowcolor[HTML]{F3F3F3} Loomio & \cellcolor{gray!25}0.2 & 0.4 & \cellcolor{gray!25} 0.5 & 0.6 & 21.9 & 3.0 & 3.1 \\
            
            \rowcolor[HTML]{FFFFFF} Languagetool... & \cellcolor{gray!25}0.3 & 0.4 & \cellcolor{gray!25} 0.5 & 0.6 & 17.3 & 1.1 & 2.2 \\
            
            \rowcolor[HTML]{F3F3F3} ocpsoft & \cellcolor{gray!25}0.2 & 0.3 & \cellcolor{gray!25} 0.3 & 0.4 & 19.3 & 1.2 & 3.3 \\
            
            \rowcolor[HTML]{FFFFFF} Locomotivems & \cellcolor{gray!25}0.2 & 0.3 & \cellcolor{gray!25} 0.3 & 0.4 & 8.4 & 1.2 & 2.2 \\
           
            \rowcolor[HTML]{F3F3F3}  Parsl & \cellcolor{gray!25}0.3 & 0.5 & \cellcolor{gray!25} 0.7 & 0.8 & 27.4 & 1.2 & 3.7 \\
            
            \rowcolor[HTML]{FFFFFF} Graylog2 & \cellcolor{gray!25}0.4 & 0.4 & \cellcolor{gray!25} 0.5 & 0.6 & 22.3 & 1.7 & 3.2 \\
            
            \rowcolor[HTML]{F3F3F3} Eclipse & \cellcolor{gray!25}0.4 & 0.6 & \cellcolor{gray!25} 0.8 & 1.0 & 86.1 & 4.0 & 8.3 \\
            
            \rowcolor[HTML]{FFFFFF} Rspec & \cellcolor{gray!25}0.5 & 0.7 & \cellcolor{gray!25} 0.8 & 1.0 & 29.8 & 7.2 & 4.0 \\
           
            \rowcolor[HTML]{F3F3F3}Radical-syber.. & \cellcolor{gray!25}0.9 & 1.3 & \cellcolor{gray!25} 1.7 & 1.9 & 69.1 & 9.7 & 6.2 \\
            
            \rowcolor[HTML]{FFFFFF} Deeplearning4j & \cellcolor{gray!25}1.0 & 1.3 & \cellcolor{gray!25} 1.6 & 1.9 & 99.0 & 7.4 & 8.8 \\
            
            \rowcolor[HTML]{F3F3F3} Puppetlabs & \cellcolor{gray!25}1.3 & 1.4 & \cellcolor{gray!25} 1.6 & 1.8 & 54.1 & 5.7 & 5.3 \\
            
            \rowcolor[HTML]{FFFFFF}Nutzam & \cellcolor{gray!25}2.2 & 3.4 & \cellcolor{gray!25} 4.5 & 5.4 & 792.7 & 17.9 & 57.4 \\
            
            \rowcolor[HTML]{F3F3F3} Square & \cellcolor{gray!25}2.7 & 3.1 & \cellcolor{gray!25} 3.9 & 4.7 & 168.5 & 11.2 & 13.1 \\
        \end{tabular} &
        \begin{tabular}{|l|r|r|r|r|r|r|r}
            \rowcolor{gray!1}
            \textbf{Project Name} & \textbf{B1} & \textbf{B2} & \textbf{B3} & \textbf{B4} & \textbf{C1} & \textbf{C2} & \textbf{D1} \\
            \hline
            \rowcolor[HTML]{F3F3F3} Middleman & \cellcolor{gray!25}3.4 & 6.2 & \cellcolor{gray!25} 8.0 & 9.4 & 598.8 & 34.5 & 33.0 \\
            
            \rowcolor[HTML]{FFFFFF} Diaspora & \cellcolor{gray!25}9.1 & 20.6 & \cellcolor{gray!25} 26.4 & 31.3 & 1738.0 & 255.8 & 78.3 \\
            
            \rowcolor[HTML]{F3F3F3}Structr & \cellcolor{gray!25}14.4 & 19.4 & \cellcolor{gray!25} 25.1 & 31.5 & 3309.9 & 129.9 & 147.9 \\
           
            \rowcolor[HTML]{FFFFFF}Yt-project & \cellcolor{gray!25} 14.7 & 27.8 & \cellcolor{gray!25} 34.6 & 39.9 & 12494.4 & 315.8 & 563.4 \\
         
            \rowcolor[HTML]{F3F3F3}Mdanalysis & \cellcolor{gray!25}24.5 & 38.1 & \cellcolor{gray!25} 47.8 & 55.5 & 34336.1 & 1271.9 & 3542.5 \\
            
            \rowcolor[HTML]{FFFFFF}Facebook & \cellcolor{gray!25}43.0 & 45.7 & \cellcolor{gray!25} 56.4 & 62.7 & 67635.8 & 281.7 & 4646.1 \\
           
            \rowcolor[HTML]{F3F3F3}Reaction.. & \cellcolor{gray!25}45.0 & 70.4 & \cellcolor{gray!25} 86.3 & 105.5 & n/a & 924.2 & 2551.4 \\
           
            \rowcolor[HTML]{FFFFFF} Openforcefield & \cellcolor{gray!25}49.2 & 81.8 & \cellcolor{gray!25} 102.4 & 117.6 & n/a & 4685.4 & 33962.7 \\
           
            \rowcolor[HTML]{F3F3F3}Unidata & \cellcolor{gray!25}105.7 & 158.5 &\cellcolor{gray!25} 196.3 & 244.9 & n/a & 1793.8 & 4271.5 \\
           
            \rowcolor[HTML]{FFFFFF} Materials.. & \cellcolor{gray!25}167.3 & 183.1 & \cellcolor{gray!25} 221.2 & 277.0 & n/a & 873.5 & 6439.4 \\
           
            \rowcolor[HTML]{F3F3F3}Spotify & \cellcolor{gray!25}564.5 & 959.8 & \cellcolor{gray!25} 1203.1 & 1541.1 & n/a & 5525.3 & 15961.1 \\
            
            \rowcolor[HTML]{FFFFFF} Rails & \cellcolor{gray!25}2950.1 & 5376.4 & \cellcolor{gray!25} 6716.6 & 8339.4 & n/a & 68648.1 & 82601.2 \\
            
            \rowcolor[HTML]{F3F3F3}Galaxy.. & \cellcolor{gray!25}9721.8 & 9169.9 & \cellcolor{gray!25} 11297.2 & 14466.7 & n/a & 56929.1 & 205651.6 \\
            
            \rowcolor[HTML]{FFFFFF}Jruby & \cellcolor{gray!25}1081.7 & 1797.7 & \cellcolor{gray!25} 2393.1 & 2946.3 & n/a & n/a & n/a \\
            
            \rowcolor[HTML]{F3F3F3} & & & & & & & \\
            
            \rowcolor[HTML]{FFFFFF}Proprietary data & 1.7 & 2.3 & 2.2 & 2.4 & 1457.5 & 285.9 & \cellcolor[HTML]{666666}{\color[HTML]{FFFFFF}327.7}
        \end{tabular}
    \end{tabular}
    \end{adjustbox}
\end{table*}

\subsection{\revised{Explanation of the Differences for the Selected Prioritization Algorithms}} \label{PerformanceofPurposedTCP}
In our study, we conduct that D1 performs ``best'' in the industrial closed-source project, but ``worst'' in open-source projects. On the opposite, B1 and B3 have the best performance in open-source projects, but worse in the industrial closed-source project.
How to explain this difference?
\\
To answer this question, we first recall the 
internal details of our different  algorithms:
\bi
    \item
    The D1 algorithm learns from the results of other test cases in the current build.  
    \item
    On the other hand, B1 and B3 are much more straight forward, since they only need the previous test results.
\ei
From the above, we would predict that D1 works best when there are more failed test cases per build in the proprietary project than our open-source projects. To check this conjecture, \revised{we count the number of} {\bf failed test cases per build}. 
\revised{Since some projects may contain thousands of builds, it is better to look at the distribution across the entire project.}
Accordingly, we report the 10th, 30th, 50th, 70th, \revised{and} 90th percentiles of those distributions. We say that our conjecture is supported if the number of failed test cases within a build is (a)~markedly different between \revised{the} open-source \revised{projects} and \revised{the proprietary project} and (b)~much larger in \revised{the} proprietary data.

\begin{figure}[t]
    \centering
    \includegraphics[width = 0.48\textwidth]{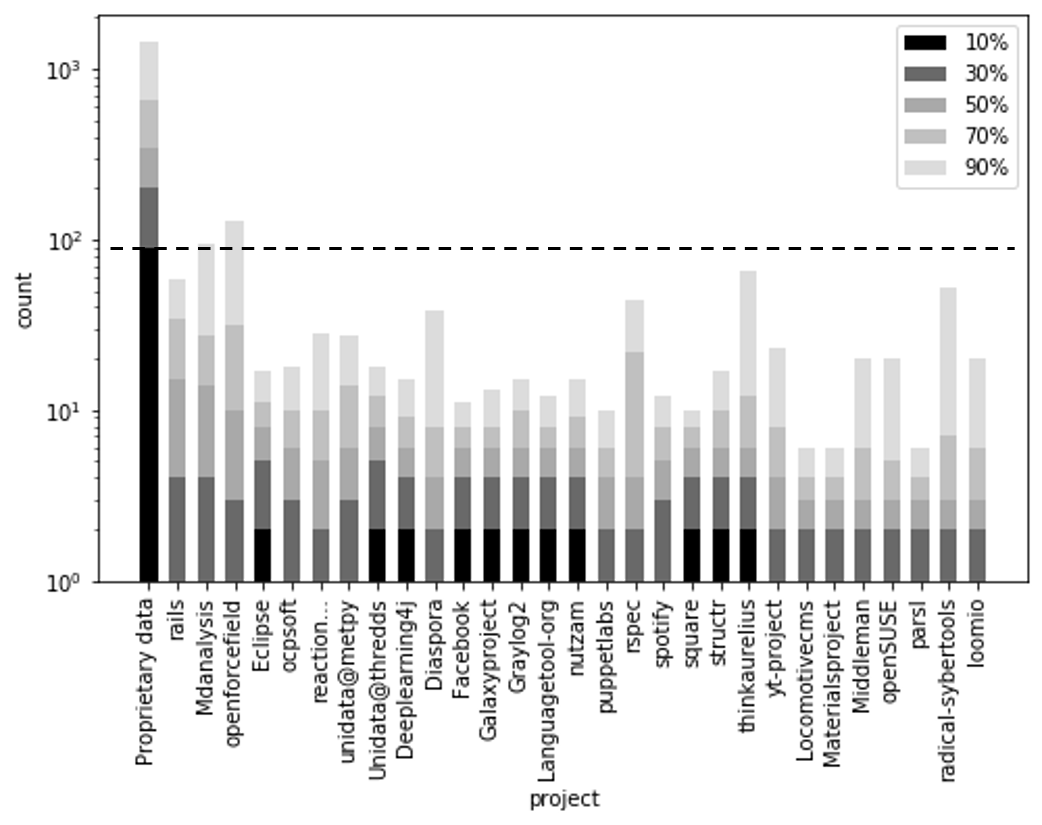}
    \caption{Figure of statistics on the number of failures in each project. The left bar shows our closed-source data while the other bars come from the open-source projects. For example, on the left-hand side, in the proprietary data, the median number of failures is 143 while on the right-hand side, in the loomio project, the median number of failures is only 1. As shown here, the worst case behavior is very different across the projects (so much so that we have to use the log scale for the y-axis). For example, the 90th percentile of the failure number (proprietary and loomio) are (791 and 14), respectively.}
    \label{fig:NumberFailure}
\end{figure}

\revised{\fig{NumberFailure} shows those results. As predicted by our conjecture, we see that the max number of failed tests per build is markedly different and larger in our proprietary than otherwise. For example, with only one exception after the log transformation (openforcefield), the 90th percentile values from open-source projects (with color {\em gainsboro}) are smaller than the 10th percentile value of our proprietary project (mark with the black dot line).}

This observation leads to the following \revised{statements} about \revised{selecting different} test case prioritization methods \revised{in different scenarios}:
\bi
    \item 
    If projects do frequent builds that address a small number of bugs each time, then there is little information in each build. In this case, we need to look more into historical data (e.g. B1 and B3).
    \item
    On the other hand, if a team tends towards ``big bang'' engineering where a release fixes multiple test failures (and, most likely, releases take longer to be generated), then there is much information in each build. In this case, an active learning approach (e.g. D1) can achieve much by reflecting over all the data in the current build.
\ei
\begin{figure}[t!]
    \centering
    \includegraphics[width=0.5\textwidth]{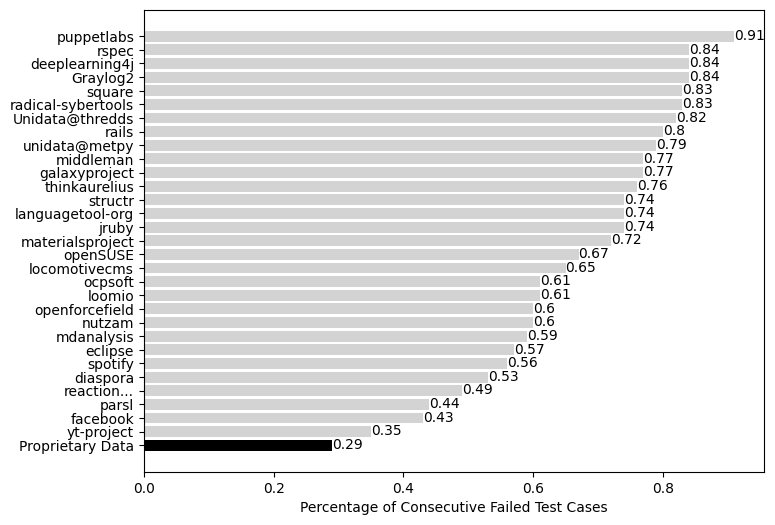}
    \caption{Percentage of consecutive failed test cases. Our proprietary closed-source project is shown last \revised{because it has the lowest percentage by comparing to other open-source projects}}
    \label{fig:ConsFailure}
\end{figure}
To further support our last point,
we make the following observation. The premise of the last point is that for ``big-bang'' projects, 
more information can be obtained by not only \revised{analyzing} \revised{the prior builds}, but also the current build.
To test that premise, we compare how much previous failing build information is available from the historical record for our proprietary and open-source projects.

\fig{ConsFailure} shows the percent of
times a failed test (from the current build) also fails in the previous build. As can be seen in that figure, in open source projects, 75\% (median) of the failing tests in the current build also failed in the last build. However, in our proprietary project, the majority (nearly three quarters) of failures in this build did \underline{{\bf not}} appear in the last build. Hence we can explain now why B1 and B3 work better in open-source than proprietary project: 
\bi
\item
B1 and B3 reflect more in the history of a project.
\item
And in that history, open-source projects have more previous build failure information.
\ei
 
Overall, the picture we are offering here is that open-source projects are more earnest about removing bugs as soon as possible. Hence:
\bi
\item
Test case prioritization methods for proprietary projects should make most use of {\em the tests from current build as well as their connection to the execution history}.
\item
While test case prioritization methods for open-source projects should only make most use of {\em the past results}.
\ei

\subsection{Efficiency of Prioritization Algorithms} \label{EfficiencyofTCPAlgorithms}
To conclude this study, we offer a brief note on the efficiency of the different prioritization
algorithms.
Efficiency can be an important component in judging whether a prioritization scheme is useful or not. An algorithm can be regarded worse than others if it takes a long duration to prioritize test cases.

In~\tbl{runtime}, we list simulation time for each of the selected algorithm. In this table, \textbf{n/a} means the algorithm takes a very long time (over 48 hours) in simulation, so we will not consider it even though its APFD is very high.

From~\tbl{runtime}, we find that our \revised{selected} algorithms B1 and B3 for open-source projects have very short execution time (marked in light gray). The reason they are fast is that they only need to analyze execution history one time for each test case (which is an $\mathcal{O}(n)$ analysis). Since most of the open-source projects have very large builds and lots of test cases, this finding consolidate our conclusion that B1 and B3 are the best prioritization algorithms in open-source projects since:
\bi
    \item
    B1 and B3 have the best performance in prioritizing open-source projects.
    \item
    B1 and B3 have fast simulation speed in prioritizing large open-source projects.
\ei
That said, despite their efficiency,  B1 and B3 are not applicable in our industrial closed-source project:
\bi
\item
In~\tbl{runtime}, we can observe that D1 can finish test case prioritization in 5 minutes with outstanding performance, which is acceptable (marked in dark gray). Although B1 and B3 only take a few seconds to finish the same task, we still prefer D1 since it can increase fault detection rates significantly.
\item Also, from~\tbl{SK_LN}, we can find B1 and B3 are only in the rank 3. D1 has a much better performance than B1 and B3. 
\ei
By taking the above reasons together, we conclude that B1 and B3 are not applicable in the closed-source project even though they have the shortest simulation time.

\section{Threats to Validity} \label{ThreatstoValidity}
This section discusses issues raised by Feldt et al.~\cite{feldt2010validity}

\textbf{Conclusion validity: }Different treatments to simulation results may cause various conclusions. In our experiments, we implement Scott-Knott analysis to the APFD results of test runs. Prioritization algorithms differ significantly if they distribute in different ranks.

\textbf{Metric validity: }We implement the weighted average of the percentage of faults detected (APFD) to evaluate the performance of prioritization approaches. This evaluation metric assumes all test cases have the same cost and the same severity. However, some test cases may take longer to execute than others. This may be a threat to evaluation validity. In our future work, we plan to collect the cost of test cases so that we can implement a better evaluation metric called the average percentage of faults detected with cost (APFDc)~\cite{elbaum2001incorporating}.

\textbf{Sampling validity: } 
One way to characterize this paper is as a response to the sampling bias problem in \revised{the} previous work. As said in the introduction,
Yu et al.~\cite{yu2019terminator} reported that the D1 TERMINATOR test case prioritization algorithm was better than dozens of alternatives. However, TERMINATOR was developed for the closed-source proprietary software. This raises the question explored here: does TERMINATOR work for other kinds of projects (e.g. open-source projects)? 
~\\
While this paper mitigates some of the sampling bias seen in  Yu et al.,
it is also true that other data, not used in this study, could invalidate our results. 
Sampling bias threatens any paper on analytics (not just this one) since conclusions that hold for one project may not hold for any other. No paper can explore all data sets -- the best we can do (and we have done) is carefully documenting our methods and placing our tools on a repository  that others can access (so the community can easily apply our methods to their data).

More specifically, this study reports results from dozens of open-source projects. While, ideally, we should also have report on an equal number of closed-source projects,  industrial SE research does not work that way. Like many other researchers, we have spend years carefully nurturing our industrial contacts (and reporting on the results of those interactions~\cite{shrikanth2019assessing,chen2019predicting,wang2019characterizing,agrawal2018we,krishna2018connection,hihn2017nasa,layman2016topic,menzies2017delayed,kocaguneli2013distributed,menzies2008learning}). Also, like many other researchers, we find it hard to get data released from industrial clients. Hence, as shown in \tbl{PaperTable}, 
researchers in this area have only been able to  use data from 0, 1 or 2 closed-source projects. 

For us,  the only closed-source data we can report here comes from the Yu et al.'s TERMINATOR study. 
Since such closed-source data is so scarce, we take care to make the best use of it:
\bi 
\item
    \S\ref{RQ3} showed that there was an unequivocal difference in results from the \revised{closed-source} data and our 30 open-source projects. Specifically, in all 30 open-source projects, the methods learned by TERMINATOR (learned from closed-source projects) failed very badly.
\item 
    In \S\ref{PerformanceofPurposedTCP}, we  showed that that difference can be explained due to  fundamental differences in the  nature of open and closed-source projects (open-source developers try to fix less bugs in consecutive builds than \revised{the close-source project developers do}). 

\ei
The lessons \revised{state} in our conclusion section are based on these two observations.



\section{Conclusion and Future Work} \label{ConclusionandFutureWork}
Regression testing is an important component in software testing and development. Better prioritization schemes can help developers detect more faults within a limited time. Therefore, test case prioritization is widely studied in the software testing region.

By searching the literature, we found nine prioritization schemes that prioritize test cases by utilizing the information of execution history. These were applied to
one closed-source project and 30 open-source projects. The differences in results
between our closed-source project and the open-source projects was clear:
\bi
\item
The D1 TERMINATOR algorithm performs the best in the industrial closed-source project but performs the worst in open-source projects. 
\item Further, algorithm B1 and B3 have the highest performance in open-source projects, while they are worse than D1 in our closed-source project.
\ei
 \S\ref{RQ3} of this paper argued that this difference was fundamental to nature
 of open and closed-source projects; specifically: 
 \bi
 \item
 Open-source developers try to fix fewer bugs in consecutive builds than close
sourced projects;
\item
This has implications on how much can be learned from one build;
\item
This, in turn, has implications on what prioritization method works best.
\ei
\tbl{PaperTable} of this report shows that this study uses far more projects than anything listed there within the last decade. Nevertheless, like many other researchers, we only have limited access to closed-source projects. Hence,
we take care to express our conclusions appropriately.
When answering  {\bf RQ3}, when recommending better prioritization schemes,
we take care to say  ``can work worse'' rather than ``will always work'
Also, we express our general conclusion 
not in terms of open-vs-closed but rather in terms of the need to tuning prioritization methods to the projects at hand

Specifically, the general lesson we offer  is:
 \begin{quote}
{\em  It is ill-advised to always apply one prioritization scheme to all projects since prioritization requires tuning to different projects types.}
\end{quote}
As to future work, we suggest the following.   It is no longer enough to report ``the'' best prioritization scheme. Research in this area should pivot to a related question; i.e. how can we, in a cost and time effective manner, explore different test case prioritization for the current data. Hence we say that future work should:
\bi
 \item
 Collect the cost of test cases from more open-source projects (to find better performance evaluation metrics).
 \item
 Make more comparisons by implementing more 
 prioritization algorithms for both open-source projects and closed-source projects.
 \item
 Collect more projects from different sources to verify our findings in both open-source projects and closed-source projects.
 \item
 Seek patterns of how the APFD score changes when the test run increases for each algorithm in the large projects. By implementing feature extraction techniques and machine learners, we can try to predict the best test case prioritization algorithm for the projects.
\item
 Develop a prioritization scheme that can work well for both open-source projects and closed-source projects.
\ei

\section*{Acknowledgements}
This work was partially funded by 
a research grant from the National Science Foundation (CCF \#1703487).

\bibliographystyle{IEEEtran}
\bibliography{bibreference}

\begin{thebibliography}{10}
\providecommand{\url}[1]{#1}
\csname url@samestyle\endcsname
\providecommand{\newblock}{\relax}
\providecommand{\bibinfo}[2]{#2}
\providecommand{\BIBentrySTDinterwordspacing}{\spaceskip=0pt\relax}
\providecommand{\BIBentryALTinterwordstretchfactor}{4}
\providecommand{\BIBentryALTinterwordspacing}{\spaceskip=\fontdimen2\font plus
\BIBentryALTinterwordstretchfactor\fontdimen3\font minus
  \fontdimen4\font\relax}
\providecommand{\BIBforeignlanguage}[2]{{%
\expandafter\ifx\csname l@#1\endcsname\relax
\typeout{** WARNING: IEEEtran.bst: No hyphenation pattern has been}%
\typeout{** loaded for the language `#1'. Using the pattern for}%
\typeout{** the default language instead.}%
\else
\language=\csname l@#1\endcsname
\fi
#2}}
\providecommand{\BIBdecl}{\relax}
\BIBdecl

\bibitem{fazlalizadeh2009prioritizing}
Y.~Fazlalizadeh, A.~Khalilian, M.~A. Azgomi, and S.~Parsa, ``Prioritizing test
  cases for resource constraint environments using historical test case
  performance data,'' in \emph{2009 2nd IEEE International Conference on
  Computer Science and Information Technology}.\hskip 1em plus 0.5em minus
  0.4em\relax IEEE, 2009, pp. 190--195.

\bibitem{lu2009introReg}
Y.~{Lu}, Y.~{Lou}, S.~{Cheng}, L.~{Zhang}, D.~{Hao}, Y.~{Zhou}, and L.~{Zhang},
  ``How does regression test prioritization perform in real-world software
  evolution?'' in \emph{2016 IEEE/ACM 38th International Conference on Software
  Engineering (ICSE)}, 2016, pp. 535--546.

\bibitem{mahajan2015intorReg}
S.~{Mahajan}, S.~D. {Joshi}, and V.~{Khanaa}, ``Component-based software system
  test case prioritization with genetic algorithm decoding technique using java
  platform,'' in \emph{2015 International Conference on Computing Communication
  Control and Automation}, 2015, pp. 847--851.

\bibitem{chittimalli2007cost}
P.~K. {Chittimalli} and M.~J. {Harrold}, ``Re-computing coverage information to
  assist regression testing,'' in \emph{2007 IEEE International Conference on
  Software Maintenance}, 2007, pp. 164--173.

\bibitem{beller2019IDE}
M.~{Beller}, G.~{Gousios}, A.~{Panichella}, S.~{Proksch}, S.~{Amann}, and
  A.~{Zaidman}, ``Developer testing in the ide: Patterns, beliefs, and
  behavior,'' \emph{IEEE Transactions on Software Engineering}, vol.~45, no.~3,
  pp. 261--284, 2019.

\bibitem{beller2015and}
M.~Beller, G.~Gousios, A.~Panichella, and A.~Zaidman, ``When, how, and why
  developers (do not) test in their ides,'' in \emph{Proceedings of the 2015
  10th Joint Meeting on Foundations of Software Engineering}, 2015, pp.
  179--190.

\bibitem{Wong1997TCPwidelyapply}
W.~E. {Wong}, J.~R. {Horgan}, S.~{London}, and H.~{Agrawal}, ``A study of
  effective regression testing in practice,'' in \emph{Proceedings The Eighth
  International Symposium on Software Reliability Engineering}, 1997, pp.
  264--274.

\bibitem{cho2016history}
Y.~Cho, J.~Kim, and E.~Lee, ``History-based test case prioritization for
  failure information,'' in \emph{2016 23rd Asia-Pacific Software Engineering
  Conference (APSEC)}.\hskip 1em plus 0.5em minus 0.4em\relax IEEE, 2016, pp.
  385--388.

\bibitem{elbaum2014techniques}
S.~Elbaum, G.~Rothermel, and J.~Penix, ``Techniques for improving regression
  testing in continuous integration development environments,'' in
  \emph{Proceedings of the 22nd ACM SIGSOFT International Symposium on
  Foundations of Software Engineering}, 2014, pp. 235--245.

\bibitem{yu2019terminator}
Z.~Yu, F.~Fahid, T.~Menzies, G.~Rothermel, K.~Patrick, and S.~Cherian,
  ``Terminator: better automated ui test case prioritization,'' in
  \emph{Proceedings of the 2019 27th ACM Joint Meeting on European Software
  Engineering Conference and Symposium on the Foundations of Software
  Engineering}, 2019, pp. 883--894.

\bibitem{koch2002effort}
S.~Koch and G.~Schneider, ``Effort, co-operation and co-ordination in an open
  source software project: Gnome,'' \emph{Information Systems Journal},
  vol.~12, no.~1, pp. 27--42, 2002.

\bibitem{raja2012defining}
U.~Raja and M.~J. Tretter, ``Defining and evaluating a measure of open source
  project survivability,'' \emph{IEEE Transactions on Software Engineering},
  vol.~38, no.~1, pp. 163--174, 2012.

\bibitem{saltis2018comparing}
S.~Saltis, ``Comparing open source software vs closed source software,''
  \emph{Core DNA}, 2018.

\bibitem{msr17challenge}
M.~Beller, G.~Gousios, and A.~Zaidman, ``Travistorrent: Synthesizing travis ci
  and github for full-stack research on continuous integration,'' in
  \emph{Proceedings of the 14th working conference on mining software
  repositories}, 2017.

\bibitem{beller2017oops}
------, ``Oops, my tests broke the build: An explorative analysis of travis ci
  with github,'' in \emph{2017 IEEE/ACM 14th International Conference on Mining
  Software Repositories (MSR)}.\hskip 1em plus 0.5em minus 0.4em\relax IEEE,
  2017, pp. 356--367.

\bibitem{rothermel2001prioritizing}
G.~Rothermel, R.~H. Untch, C.~Chu, and M.~J. Harrold, ``Prioritizing test cases
  for regression testing,'' \emph{IEEE Transactions on software engineering},
  vol.~27, no.~10, pp. 929--948, 2001.

\bibitem{elbaum2002test}
S.~Elbaum, A.~G. Malishevsky, and G.~Rothermel, ``Test case prioritization: A
  family of empirical studies,'' \emph{IEEE transactions on software
  engineering}, vol.~28, no.~2, pp. 159--182, 2002.

\bibitem{li2007search}
Z.~Li, M.~Harman, and R.~M. Hierons, ``Search algorithms for regression test
  case prioritization,'' \emph{IEEE Transactions on software engineering},
  vol.~33, no.~4, pp. 225--237, 2007.

\bibitem{kim2002history}
J.-M. Kim and A.~Porter, ``A history-based test prioritization technique for
  regression testing in resource constrained environments,'' in
  \emph{Proceedings of the 24th international conference on software
  engineering}, 2002, pp. 119--129.

\bibitem{jiang2009adaptive}
B.~Jiang, Z.~Zhang, W.~K. Chan, and T.~Tse, ``Adaptive random test case
  prioritization,'' in \emph{2009 IEEE/ACM International Conference on
  Automated Software Engineering}.\hskip 1em plus 0.5em minus 0.4em\relax IEEE,
  2009, pp. 233--244.

\bibitem{srikanth2005system}
H.~Srikanth, L.~Williams, and J.~Osborne, ``System test case prioritization of
  new and regression test cases,'' in \emph{2005 International Symposium on
  Empirical Software Engineering, 2005.}\hskip 1em plus 0.5em minus 0.4em\relax
  IEEE, 2005, pp. 10--pp.

\bibitem{carlson2011clustering}
R.~Carlson, H.~Do, and A.~Denton, ``A clustering approach to improving test
  case prioritization: An industrial case study,'' in \emph{2011 27th IEEE
  International Conference on Software Maintenance (ICSM)}.\hskip 1em plus
  0.5em minus 0.4em\relax IEEE, 2011, pp. 382--391.

\bibitem{qu2007test}
B.~Qu, C.~Nie, B.~Xu, and X.~Zhang, ``Test case prioritization for black box
  testing,'' in \emph{31st Annual International Computer Software and
  Applications Conference (COMPSAC 2007)}, vol.~1.\hskip 1em plus 0.5em minus
  0.4em\relax IEEE, 2007, pp. 465--474.

\bibitem{marijan2013TCP}
D.~{Marijan}, A.~{Gotlieb}, and S.~{Sen}, ``Test case prioritization for
  continuous regression testing: An industrial case study,'' in \emph{2013 IEEE
  International Conference on Software Maintenance}, 2013, pp. 540--543.

\bibitem{hemmati2015prioritizing}
H.~Hemmati, Z.~Fang, and M.~V. Mantyla, ``Prioritizing manual test cases in
  traditional and rapid release environments,'' in \emph{2015 IEEE 8th
  International Conference on Software Testing, Verification and Validation
  (ICST)}.\hskip 1em plus 0.5em minus 0.4em\relax IEEE, 2015, pp. 1--10.

\bibitem{zhu2018test}
Y.~Zhu, E.~Shihab, and P.~C. Rigby, ``Test re-prioritization in continuous
  testing environments,'' in \emph{2018 IEEE International Conference on
  Software Maintenance and Evolution (ICSME)}.\hskip 1em plus 0.5em minus
  0.4em\relax IEEE, 2018, pp. 69--79.

\bibitem{rothermel1999test}
G.~Rothermel, R.~H. Untch, C.~Chu, and M.~J. Harrold, ``Test case
  prioritization: An empirical study,'' in \emph{Proceedings IEEE International
  Conference on Software Maintenance-1999 (ICSM'99).'Software Maintenance for
  Business Change'(Cat. No. 99CB36360)}.\hskip 1em plus 0.5em minus 0.4em\relax
  IEEE, 1999, pp. 179--188.

\bibitem{elbaum2001incorporating}
S.~Elbaum, A.~Malishevsky, and G.~Rothermel, ``Incorporating varying test costs
  and fault severities into test case prioritization,'' in \emph{Proceedings of
  the 23rd International Conference on Software Engineering. ICSE 2001}.\hskip
  1em plus 0.5em minus 0.4em\relax IEEE, 2001, pp. 329--338.

\bibitem{do2006use}
H.~Do and G.~Rothermel, ``On the use of mutation faults in empirical
  assessments of test case prioritization techniques,'' \emph{IEEE Transactions
  on Software Engineering}, vol.~32, no.~9, pp. 733--752, 2006.

\bibitem{haidry2013TCP}
S.~{Haidry} and T.~{Miller}, ``Using dependency structures for prioritization
  of functional test suites,'' \emph{IEEE Transactions on Software
  Engineering}, vol.~39, no.~2, pp. 258--275, 2013.

\bibitem{Malz2012TCP}
C.~{Malz}, N.~{Jazdi}, and P.~{Gohner}, ``Prioritization of test cases using
  software agents and fuzzy logic,'' in \emph{2012 IEEE Fifth International
  Conference on Software Testing, Verification and Validation}, 2012, pp.
  483--486.

\bibitem{zemlin2017}
J.~Zemlin, ``If you can’t measure it, you can’t improve it. https://www.
  linux.com/news/if-you-cant-measure-it-you-cant-improve-it-chaoss-project-creates-
  tools-analyze-software/,'' 2017.

\bibitem{vance2010legal}
A.~Vance, ``Legal sites plan revamps as rivals undercut price,'' \emph{The New
  York Times}, 2010.

\bibitem{elbaum2004selecting}
S.~Elbaum, G.~Rothermel, S.~Kanduri, and A.~G. Malishevsky, ``Selecting a
  cost-effective test case prioritization technique,'' \emph{Software Quality
  Journal}, vol.~12, no.~3, pp. 185--210, 2004.

\bibitem{malishevsky2006cost}
A.~G. Malishevsky, J.~R. Ruthruff, G.~Rothermel, and S.~Elbaum,
  ``Cost-cognizant test case prioritization,'' Technical Report
  TR-UNL-CSE-2006-0004, University of Nebraska-Lincoln, Tech. Rep., 2006.

\bibitem{zhang2009time}
L.~Zhang, S.-S. Hou, C.~Guo, T.~Xie, and H.~Mei, ``Time-aware test-case
  prioritization using integer linear programming,'' in \emph{Proceedings of
  the eighteenth international symposium on Software testing and analysis},
  2009, pp. 213--224.

\bibitem{jeffrey2006test}
D.~Jeffrey and N.~Gupta, ``Test case prioritization using relevant slices,'' in
  \emph{30th Annual International Computer Software and Applications Conference
  (COMPSAC'06)}, vol.~1.\hskip 1em plus 0.5em minus 0.4em\relax IEEE, 2006, pp.
  411--420.

\bibitem{tu2020changing}
H.~Tu, R.~Agrawal, and T.~Menzies, ``The changing nature of computational
  science software,'' \emph{arXiv preprint arXiv:2003.05922}, 2020.

\bibitem{kalliamvakou2014promises}
E.~Kalliamvakou, G.~Gousios, K.~Blincoe, L.~Singer, D.~M. German, and
  D.~Damian, ``The promises and perils of mining github,'' in \emph{Proceedings
  of the 11th working conference on mining software repositories}, 2014, pp.
  92--101.

\bibitem{munaiah2017curating}
N.~Munaiah, S.~Kroh, C.~Cabrey, and M.~Nagappan, ``Curating github for
  engineered software projects,'' \emph{Empirical Software Engineering},
  vol.~22, no.~6, pp. 3219--3253, 2017.

\bibitem{lnnyt1}
\BIBentryALTinterwordspacing
``Company news; a name change is planned for mead data central,'' \emph{The New
  York Times.}, 1994. [Online]. Available:
  \url{https://www.nytimes.com/1994/12/02/business/company-news-a-name-change-is-planned-for-mead-data-central.html?src=pm}
\BIBentrySTDinterwordspacing

\bibitem{lnnyt}
\BIBentryALTinterwordspacing
A.~Vance, ``Legal sites plan revamps as rivals undercut price,'' \emph{The New
  York Times.}, 2010. [Online]. Available:
  \url{https://www.nytimes.com/2010/01/25/technology/25westlaw.html?_r=1&ref=reedelsevier}
\BIBentrySTDinterwordspacing

\bibitem{lnnyt2}
\BIBentryALTinterwordspacing
S.~Miller, ``For future reference, a pioneer in online reading,'' \emph{The
  Wall Street Journal}, 2012. [Online]. Available:
  \url{https://www.wsj.com/articles/SB10001424052970203721704577157211501855648?KEYWORDS=lexisnexis}
\BIBentrySTDinterwordspacing

\bibitem{mittas2012ranking}
N.~Mittas and L.~Angelis, ``Ranking and clustering software cost estimation
  models through a multiple comparisons algorithm,'' \emph{IEEE Transactions on
  software engineering}, vol.~39, no.~4, pp. 537--551, 2012.

\bibitem{xia2018hyperparameter}
T.~Xia, R.~Krishna, J.~Chen, G.~Mathew, X.~Shen, and T.~Menzies,
  ``Hyperparameter optimization for effort estimation,'' \emph{arXiv preprint
  arXiv:1805.00336}, 2018.

\bibitem{macbeth2011cliff}
G.~Macbeth, E.~Razumiejczyk, and R.~D. Ledesma, ``Cliff's delta calculator: A
  non-parametric effect size program for two groups of observations,''
  \emph{Universitas Psychologica}, vol.~10, no.~2, pp. 545--555, 2011.

\bibitem{hess2004robust}
M.~R. Hess and J.~D. Kromrey, ``Robust confidence intervals for effect sizes: A
  comparative study of cohen’sd and cliff’s delta under non-normality and
  heterogeneous variances,'' in \emph{annual meeting of the American
  Educational Research Association}, 2004, pp. 1--30.

\bibitem{feldt2010validity}
R.~Feldt and A.~Magazinius, ``Validity threats in empirical software
  engineering research-an initial survey.'' in \emph{Seke}, 2010, pp. 374--379.

\bibitem{shrikanth2019assessing}
N.~Shrikanth and T.~Menzies, ``Assessing practitioner beliefs,'' \emph{arXiv
  preprint arXiv:1912.10093}, 2019.

\bibitem{chen2019predicting}
J.~Chen, J.~Chakraborty, P.~Clark, K.~Haverlock, S.~Cherian, and T.~Menzies,
  ``Predicting breakdowns in cloud services (with spike),'' in
  \emph{Proceedings of the 2019 27th ACM Joint Meeting on European Software
  Engineering Conference and Symposium on the Foundations of Software
  Engineering}, 2019, pp. 916--924.

\bibitem{wang2019characterizing}
J.~Wang, S.~Wang, J.~Chen, T.~Menzies, Q.~Cui, M.~Xie, and Q.~Wang,
  ``Characterizing crowds to better optimize worker recommendation in
  crowdsourced testing,'' \emph{IEEE Transactions on Software Engineering},
  2019.

\bibitem{agrawal2018we}
A.~Agrawal, A.~Rahman, R.~Krishna, A.~Sobran, and T.~Menzies, ``We don't need
  another hero? the impact of" heroes" on software development,'' in
  \emph{Proceedings of the 40th International Conference on Software
  Engineering: Software Engineering in Practice}, 2018, pp. 245--253.

\bibitem{krishna2018connection}
R.~Krishna, A.~Agrawal, A.~Rahman, A.~Sobran, and T.~Menzies, ``What is the
  connection between issues, bugs, and enhancements?'' in \emph{2018 IEEE/ACM
  40th International Conference on Software Engineering: Software Engineering
  in Practice Track (ICSE-SEIP)}.\hskip 1em plus 0.5em minus 0.4em\relax IEEE,
  2018, pp. 306--315.

\bibitem{hihn2017nasa}
J.~Hihn, M.~Saing, E.~Huntington, J.~Johnson, T.~Menzies, and G.~Mathew, ``The
  nasa analogy software cost model: A web-based cost analysis tool,'' in
  \emph{2017 IEEE Aerospace Conference}.\hskip 1em plus 0.5em minus 0.4em\relax
  IEEE, 2017, pp. 1--17.

\bibitem{layman2016topic}
L.~Layman, A.~P. Nikora, J.~Meek, and T.~Menzies, ``Topic modeling of nasa
  space system problem reports: research in practice,'' in \emph{Proceedings of
  the 13th International Conference on Mining Software Repositories}, 2016, pp.
  303--314.

\bibitem{menzies2017delayed}
T.~Menzies, W.~Nichols, F.~Shull, and L.~Layman, ``Are delayed issues harder to
  resolve? revisiting cost-to-fix of defects throughout the lifecycle,''
  \emph{Empirical Software Engineering}, vol.~22, no.~4, pp. 1903--1935, 2017.

\bibitem{kocaguneli2013distributed}
E.~Kocaguneli, T.~Zimmermann, C.~Bird, N.~Nagappan, and T.~Menzies,
  ``Distributed development considered harmful?'' in \emph{2013 35th
  International Conference on Software Engineering (ICSE)}.\hskip 1em plus
  0.5em minus 0.4em\relax IEEE, 2013, pp. 882--890.

\bibitem{menzies2008learning}
T.~Menzies, M.~Benson, K.~Costello, C.~Moats, M.~Northey, and J.~Richardson,
  ``Learning better iv\&v practices,'' \emph{Innovations in Systems and
  Software Engineering}, vol.~4, no.~2, pp. 169--183, 2008.

\end{thebibliography}

\newpage
\begin{IEEEbiography}[{\includegraphics[width=1.05in,clip,keepaspectratio]{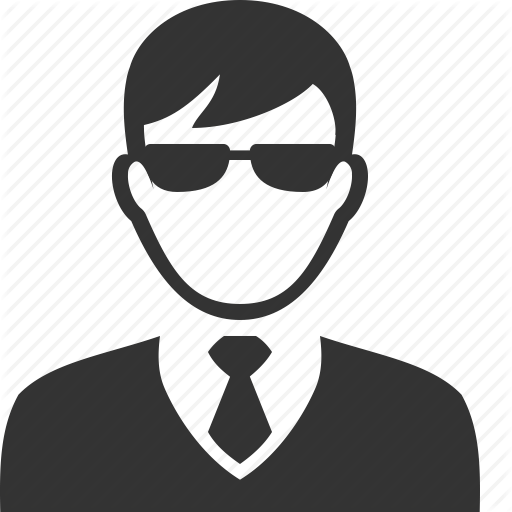}}]{Xiao Ling} is a first-year PhD student in Computer Science at NC State University. His research interests include automated software testing and machine learning for software engineering.
\end{IEEEbiography}

\begin{IEEEbiography}[{\includegraphics[width=1.05in,clip,keepaspectratio]{secret_agent.png}}]{Rishabh Agrawal} is a masters student in computer science department at NC State University. His research interests include machine learning for software engineering, data mining and deep learning.
\end{IEEEbiography}

\begin{IEEEbiography}[{\includegraphics[width=1.05in,clip,keepaspectratio]{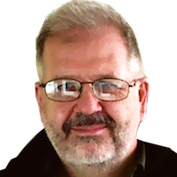}}]{Tim Menzies} (IEEE Fellow, Ph.D. UNSW, 1995)
is a Professor in computer science  at NC State University, USA,  
where he teaches software engineering,
automated software engineering,
and programming languages.
His research interests include software engineering (SE), data mining, artificial intelligence, and search-based SE, open access science. 
For more information,  please visit \url{http://menzies.us}.
\end{IEEEbiography}

\end{document}